\documentclass[twocolumn]{aastex631}

\usepackage{CJK}
\usepackage{graphicx}
\usepackage{dcolumn}
\usepackage{bm}
\usepackage{amssymb}
\usepackage{latexsym}
\usepackage{booktabs}
\usepackage{amsmath}
\newcommand{\be}{\begin{equation}}
\newcommand{\ee}{\end{equation}}
\newcommand{\bq}{\begin{eqnarray}}
\newcommand{\eq}{\end{eqnarray}}
\newcommand{\RNum}[1]{\uppercase\expandafter{\romannumeral #1\relax}}

\newcommand{\n}{\nonumber}
\def\({\left(}
\def\){\right)}
\def\[{\left[}
\def\]{\right]}
\newcommand{\Mpc}{{\,\rm Mpc}}
\newcommand{\MHz}{{\,\rm MHz}}
\newcommand{\mK}{{\,\rm mK}}

\begin{document}
\begin{CJK*}{UTF8}{gbsn}

\title{Effects of Small-Scale Absorption Systems on Neutral Islands during the Late Epoch of Reionization}
\author[0000-0002-0113-9499]{Peng-Ju Wu (武鹏举)}
\affiliation{Department of Physics, College of Sciences, Northeastern University, Shenyang 110819, China}

\author[0000-0003-3224-4125]{Yidong Xu (徐怡冬)}
\affiliation{National Astronomical Observatories, Chinese Academy of Sciences, Beijing 100101, China}

\author[0000-0002-6029-1933]{Xin Zhang (张鑫)}
\affiliation{Department of Physics, College of Sciences, Northeastern University, Shenyang 110819, China}
\affiliation{Key Laboratory of Data Analytics and Optimization for Smart Industry (Ministry of Education), Northeastern University, Shenyang 110819, China}
\affiliation{Center for High Energy Physics, Peking University, Beijing 100871, China}

\author[0000-0001-6475-8863]{Xuelei Chen (陈学雷)}
\affiliation{Department of Physics, College of Sciences, Northeastern University, Shenyang 110819, China}
\affiliation{National Astronomical Observatories, Chinese Academy of Sciences, Beijing 100101, China}
\affiliation{University of Chinese Academy of Sciences, Beijing 100049, China}
\affiliation{Center for High Energy Physics, Peking University, Beijing 100871, China}

\email{xuyd@nao.cas.cn (YX), zhangxin@mail.neu.edu.cn (XZ), xuelei@cosmology.bao.ac.cn (XC)}

\begin{abstract}

The reionization process is expected to be prolonged by the small-scale absorbers (SSAs) of ionizing photons, which have been seen as Lyman-limit systems in quasar absorption line observations. We use a set of semi-numerical simulations to investigate the effects of absorption systems on the reionization process, especially their impacts on the neutral islands during the late Epoch of Reionization (EoR). Three models are studied, i.e. the extreme case of no-SSA model with a high level of ionizing background, the moderate-SSA model with a relatively high level of ionizing background, and the dense-SSA model with a low level of ionizing background. We find that while the characteristic scale of neutral regions decreases during the early and middle stages of reionization, it stays nearly unchanged at about 10 comoving Mpc during the late stage for the no-SSA and moderate-SSA models. However, in the case of weak ionizing background in the dense-SSA model, the characteristic island scale shows obvious evolution, as large islands break into many small ones that are slowly ionized.  The evolutionary behavior of neutral islands during the late EoR thus provides a novel way to constrain the abundance of SSAs. We discuss the 21-cm observation with the upcoming Square Kilometre Array. The different models can be distinguished by either the 21-cm imaging or the 21-cm power spectrum measurements.
\end{abstract}


\section{Introduction}

Cosmic reionization is a significant phase transition in the global history of the Universe, during which the hydrogen atoms in the intergalactic medium (IGM) are ionized by photons escaped from the galaxies. Recent observations of the cosmic microwave background (CMB) and the high-redshift Lyman-$\alpha$ absorptions indicate a relatively late completion of the reionization process (e.g. \citealt{Planck2020,Becker2015,Qin2021Lya,Becker2021,Bosman2021}), with potential neutral islands lasting up to below redshift 6 \citep{Kulkarni2019,Keating2020}. Increasing attention has been driven to the late EoR, investigating the statistics of the large-scale neutral regions (islands) \citep{Giri2019,Nasir2020}, how the islands evolves \citep{Xu2014,Xu2017}, and a statistical tool has been proposed to determine the completing redshift of reionization by measuring the 21-cm bias \citep{XuWX2019}.

The 21-cm transition of neutral hydrogen (\textsc{H\,i}) has the potential to trace the whole history of cosmic dawn and reionization (see e.g. \citealt{FOB2006} for a review). The Experiment to Detect the Global Epoch of Reionization Signature (EDGES) has detected an absorption feature in the the global radio spectrum which could be the cosmic dawn signature \citep{Bowman2018}. On the other hand, various low-frequency interferometers have put increasingly lower upper limit on the 21-cm power spectrum from the EoR (e.g. \citealt{TrottMWA2020,MertensLOFAR2020,HERA2021arXiv210802263T}). The upcoming Square Kilometre Array (SKA) will have the ability to do tomographic imaging of the whole reionization process \citep{KoopmansSKA2015,2020SCPMA..6370431X}.

During the late EoR, when the ionized regions become overlapped and inter-connected with each other, the characteristic scale of ionized regions exceeds the mean free path (MFP) of the ionizing photons. Besides the large scale neutral islands, the propagation of the ionizing photons are significantly affected by small-scale absorbers (SSAs) \citep{2016ApJ...831...86P}. These absorbers have also been observed in high redshift quasar absorption lines, in particular the numerous Lyman-limit systems (LLSs) which have sufficient column density to remain neutral.  In the following, we will refer to all of these as SSAs. These played a significant role in regulating the ionizing background and the reionization process (e.g. \citealt{MHR00,Alvarez2012,Xu2017}). Therefore, it is essential to properly model the effects of SSAs when predicting and interpreting the upcoming 21-cm data, especially for the signal from the late EoR (e.g. \citealt{Sobacchi2014}).

However, the SSAs are currently not resolvable in large-scale ($> 100 \Mpc$) reionization simulations. They are usually implemented with approximations, either by implementing a sub-grid clumping (see, e.g. \citet{Iliev2007} for full-numerical simulations, and \citet{Sobacchi2014} for semi-numerical simulations), or by introducing a MFP parameter (see, e.g. \citet{Iliev2014,Shukla2016} for full-numerical simulations, and \citet{Alvarez2012,Xu2017} for semi-numerical simulations). In particular, \citet{Alvarez2012} used a semi-numerical approach to investigate the effect of the abundance of SSAs on the large-scale morphology and the overall progress of reionization, by varying the MFP parameter, and found that the characteristic bubble size is quite sensitive to the MFP of ionizing photons. \citet{Sobacchi2014} and \citet{Shukla2016} used different approaches, but arrived at the same conclusions that the presence of SSAs impeded the late growth of the ionized regions and suppressed the 21-cm power spectrum on large scales substantially.

These previous studies are primarily focused on the ionized regions, and the evolution of neutral islands are less discussed. One exception is \citet{Giri2019}, in which a comprehensive study of the statistical properties of neutral islands was presented. They showed different topology and evolutionary features between the ionized bubbles in the early EoR and the neutral islands in the late EoR. Being fully numerical, however, it has been a challenge to explore a large parameter space,
and the effect of SSAs on the island statistics is not well studied.

In the present work, we focus on the island stage, i.e. the last stage in the topological evolution of reionization \citep{ChenMF2019}, when the SSAs are most important in regulating the reionization process. We use semi-numerical simulations to explore a range of parameters that are consistent with a variety of current observations, i.e. the latest Thompson optical depth to the CMB inferred by the Planck data \citep{Planck2020}, the ionization state of the IGM inferred from high-redshift galaxy and quasar observations (e.g. \citealt{McGreer2015,Mesinger2015,Sobacchi2015,Davies2018,Mason2018,Mason2019,2020ApJ...904..144J}), the observed number density of LLSs at high redshifts (e.g. \citealt{Songaila2010};  \citealt{Crighton2019}), and measurements of the high-redshift \textsc{H\,i} photo-ionization rate (e.g. \citealt{Wyithe2011,Calverley2011,D'Aloisio2018}). We investigate the effect of SSAs on the morphology and size evolution of the neutral islands,  discuss the degeneracies between the abundance of SSAs and the properties of ionizing sources, and potential ways to break them, and explore the observational strategy to distinguish the different models in upcoming 21-cm tomographic observations.

Various techniques have been developed to measure the MFP of ionizing photons, such as using the incidence rate of LLSs (e.g. \citealt{Meiksin1993,Songaila2010}), or directly measuring the opacity from the mean spectra of high redshift quasars (e.g. \citealt{Prochaska2009,Worseck2014}), or averaging the free paths along multiple lines of sight against quasars \citep{Romano2019}.
With the increasing sample of high redshift quasars, measurements of the abundance of LLSs and the MFP have effectively
constrained the reionization process (e.g. \citealt{Cain2021,Davies2021}). However, it is still challenging to measure precisely the
MFP in the reionization epoch. Here we propose that the 21-cm observations can
be a novel probe to constrain the abundance of SSAs and the MFP of ionizing photons during the EoR.

This paper is organized as follows. We briefly describe the semi-numerical simulations used in our study and the models for SSAs in Section \ref{simulation}. Section \ref{results} contains the theoretical results on the effects of SSAs on the neutral islands, in terms of the size distribution and evolution of islands, and the power spectrum of the 21-cm brightness emitted by the islands. We also discuss the long troughs observed in quasar absorption spectrum.
In Section \ref{ska}, we compare the 21-cm imaging observation and the power spectrum measurement, and discuss the prospects of distinguishing different reionization models with the low-frequency array of phase one SKA (SKA1-Low). And we conclude in Section \ref{summary}. Throughout this paper, we assume the standard $\Lambda$CDM model with the cosmological parameters of $\Omega_{\Lambda} = 0.685$, $\Omega_{\rm m} = 0.315$, $\Omega_{\rm b} = 0.0495$, $H_0 = 67.4\ \rm{km\ s^{-1}\ {Mpc}^{-1}}$, $\sigma_8 = 0.811$, and $n_{\rm s} = 0.965$.

\section{The Semi-numerical Simulations}\label{simulation}
In order to investigate the effect of SSAs on the neutral islands, we use the semi-numerical simulation {\tt islandFAST} \citep{Xu2017} which is based on the island model \citep{Xu2014} and was developed for the isolated islands topology during the late EoR. Being semi-numerical, it is more efficient in exploring the parameter space, so that we can study the effects within the parameter space constrained by current observations.

In the framework of the excursion-set theory \citep{Bond1991,Lacey1993,FZH2004}, the {\tt islandFAST} filters the evolved density fields over increasingly smaller scales, and determines each cell in the simulation box as ionized (or neutral) if the number of ionizing photons available within the filtering scale, corrected for recombinations, exceeds (or drops below) the number of hydrogen atoms in this region. This ionization/neutral criterion is called a barrier in the excursion set theory. For the early EoR till the percolation of ionized regions , the {\tt islandFAST} is identical to the original version of {\tt 21cmFAST} \citep{21CMFAST2011}, in which the inhomogeneous recombination \citep{Sobacchi2014} and minihalos \citep{Qin2020minihalo,Qin2021minihalo} are not yet incorporated, so as to be consistent with the homogeneous ionizing background implemented for the late EoR. Assuming that the number of the ionizing photons emitted is proportional to the total collapse fraction of the region, the ionization condition, i.e. the bubble barrier can be written as
\begin{equation}
f_{\rm coll} \geq {\xi}^{-1},
\label{eq:bubble_barrier}
\end{equation}
where ${\xi} = f_{\rm esc}f_{\star}N_{\gamma / \rm H}(1+{\bar{n}_{\rm rec}})^{-1}$ is the ionizing efficiency parameter, in which $f_{\rm esc}$, $f_{\star}$, $N_{\gamma / \rm H}$ and ${\bar{n}_{\rm rec}}$ are the escape fraction, star-formation efficiency, the number of ionizing photons emitted per H atom in stars, and the average number of recombinations per ionized hydrogen atom, respectively.

For the island stage after percolation, when the neutral islands are isolated, the {\tt islandFAST} adopts an inverse topology, and the condition for a region of mass scale $M$ to keep from being fully-ionized, i.e. the island barrier, is written as
\begin{equation}
\xi\, f_{\rm coll}(\delta_{\rm M}; M,z) +\frac{\Omega_{\rm m}}{\Omega_{\rm b}} \frac{N_{\rm back}\, m_{\rm H}}{M\,X_{\rm H}\,(1+\bar{n}_{\rm rec})} < 1,
\label{eq:island_barrier}
\end{equation}
where $\delta_{\rm M}$ is the mean overdensity of the region under consideration, and $X_{\rm H}$ is the mass fraction
of baryons in hydrogen. Here $N_{\rm back}$ is the number of background ionizing photons consumed by the region
under consideration, which can be written as
\begin{equation}
N_{\rm back} = \frac{4\,\pi}{3}\, (R_i^3 - R_f^3)\, \bar{n}_{\rm b}\, (1+\bar{n}_{\rm rec}),
\end{equation}
where $\bar{n}_{\rm b}$ is the mean comoving baryon number density, $R_i$ is the initial island scale corresponding to the mass
scale $M$ when the ionizing background is just set up, and $R_f$ is final scale of the island at the redshift under consideration.
The change in the island scale is due to the ionization by the ionizing background, then we have
\begin{equation}\label{Eq.DeltaR}
R_i - R_f = \int_z^{z_{\rm back}} \frac{F(z)}{\bar{n}_{\rm b}(1+\bar{n}_{\rm rec})}
\frac{{\rm d}z}{H(z)(1+z)^3},
\end{equation}
where $F(z)$ is the physical photon number flux of the ionizing background, $z_{\rm back}$ is the redshift at which a global ionizing background is set up, and $H(z)$ is the Hubble parameter.
The number density of background ionizing photons can be written as \citep{Xu2017}
\begin{eqnarray}\label{Eq.n_gamma}
n_{\gamma}(z) &=& \int_z\, \bar{n}_{\rm H}\, \left| \frac{{\rm d}f_{\rm coll}(z')}{{\rm d}z'}\right|\, f_\star\, N_{\gamma / \rm H}\, f_{\rm esc}
\nonumber \\
&\times& \exp\left[- \frac{l(z,z')}{\lambda_{\rm mfp}(z)} \right] (1-f^{\rm host}_{\textsc{H\,i}})\, {\rm d}z',
\end{eqnarray}
where $l(z,z')$ is the physical distance between the source at $z'$ and the redshift $z$ under consideration, and $f^{\rm host}_{\textsc{H\,i}}$ is the volume fraction of host islands.
The ionizing photon number flux $F(z)$ in Eq.~(\ref{Eq.DeltaR}) is then computed with the relation $F(z) = n_{\gamma}(z)(1 + z)^3 c/4$.
Note that in {\tt islandFAST} the MFP is implemented by its original definition of attenuating ionizing photons, but not a sharp boundary for the filtering scale, or a sharp cut-off for the traveling distance of ionizing photons, as in previous simulations (e.g. \citealt{Alvarez2012,Iliev2014,Shukla2016}).
In this work, ${\xi}$ is defined in terms of $N_{\gamma / \rm H}$, and we assume the same production rate of
ionizing photons per He atom in stars as that per H atom in stars, i.e. $N_{\gamma / {\rm He}} = N_{\gamma / \rm H}$,
so Eq. (\ref{eq:island_barrier}) implicitly includes the contribution and
consumption of ionizing photons by helium atoms, and the factor $m_{\rm H}/X_{\rm H}$ in the second term in
Eq. (\ref{eq:island_barrier}) cancels out the same factor when computing $\bar{n}_{\rm H}$ in Eq.(\ref{Eq.n_gamma}).

For the island stage, the {\tt islandFAST} takes a two-step filtering approach; it first finds host islands by identifying first-down-crossings with respect to the island barrier including the contribution from an ionizing background, and then identifies bubbles-in-island by applying a second filtering step within each host islands using the bubble barrier. With an input model of SSAs, the {\tt islandFAST} self-consistently computes the effective MFP limited by both the large-scale neutral islands and the SSAs, and derive the evolution of the ionizing background simultaneously with the ionization field in adaptive redshift steps.

In the original version of {\tt islandFAST} \citep{Xu2017}, the ionizing background was assumed to be present only in the island stage. However, the percolation process begins as early as when the Universe was only $\sim 30\%$ ionized \citep{2016MNRAS.457.1813F,ChenMF2019}, and the ionizing background is expected to gradually set up since then, though it may fluctuates significantly during the percolation process. In this work, we assume that a relatively uniform ionizing background is present since the ``neutral fibers'' stage (i.e. when the neutral regions thinned into filamentary structures) \citep{ChenMF2019}, and
$z_{\rm back}$ is set by the time when the mean neutral fraction is 30\%.

We incorporate three different models for the SSAs, as detailed below. As shown in previous works, more abundant SSAs result in more delayed and prolonged reionization process (e.g. \citealt{Alvarez2012,Sobacchi2014}). In addition to the SSA models, two key parameters determine the process of reionization, i.e. the minimum mass $M_{\rm min}$, or equivalently the minimum virial temperature $T^{\rm vir}_{\rm min}$, of halos that contribute ionizing photons, and the ionizing efficiency parameter $\xi$. The former regulates the time at which the reionization begins in earnest, while the latter determines the speed of reionization process. By tuning these two parameters, we control the reionization processes with different SSA models to be consistent with various observations.

\subsection{The models for SSAs} 
During the island stage, the MFP of ionizing photons is limited not only by large-scale under-dense islands but also by small-scale over-dense absorbers. 
The effective MFP of ionizing photons $\lambda_{\rm mfp}$ is given by
\begin{equation}
\lambda_{\rm mfp}^{-1} = \lambda_{\rm I}^{-1} + \lambda_{\rm abs}^{-1},
\end{equation}
where $\lambda_{\rm I}$ is the MFP limited by large-scale under-dense islands, and $ \lambda_{\rm abs}$ is the MFP due to small-scale over-dense absorbers.
In the simulation, $\lambda_{\rm I}$ is computed on-the-fly using the mean-free-path algorithm \citep{Mesinger2007},
which is taken as the average length of random vectors starting from an ionized pixel reaching an edge of a neutral island.
 Because of the relatively low column density of Ly$\alpha$ forest systems and the relative rarity of DLAs, the LLSs are the dominant contributor to the IGM opacity, reducing the MFP of ionizing photons, and weakening the ionizing background \citep{FOtaxing2005,Shukla2016}. Here the parameter $\lambda_{\rm abs}$ implicitly includes the contributions from all kinds of unresolved absorbers, i.e. the SSAs.

We consider the following three models of SSAs corresponding to different levels of the ionizing background.

\begin{description}
\item[islandFAST-noSSA]
In the extreme case of no SSA, the MFP of ionizing photons is completely determined by the morphology of large-scale neutral islands, i.e.
$\lambda_{\rm mfp} = \lambda_{\rm I}$, the ionization background is large in this case. Though unrealistic, we can use this case for comparison,

\item[islandFAST-SC]
The number density of LLSs was observed up to redshift $\sim 6$ \citep{Songaila2010}. In the second model, we adopt the evolution of the MFP limited by SSAs extrapolated from the fitting formula provided by \citet{Songaila2010}, expressed as:
\begin{equation}\label{eq.mfp_SC}
\lambda_{\rm abs} = 50\left(\displaystyle{\frac{1+z}{4.5}}\right)^{-4.44}\ [\rm{physical\; Mpc}],
\end{equation}
This model corresponds to a relatively high level of the ionizing background, or equivalently a moderate MFP.

\item[islandFAST-RS]
Recent observations have favored a rapid evolution of the MFP near the end of reionization (e.g. \citealt{Becker2021}), implying significant modulation of the MFP by the large-scale ionization field. The {\tt islandFAST-RS} model adopts the neutral-fraction-dependent ${\lambda_{\rm mfp}}$ given by the high-resolution Aurora radiation-hydrodynamical simulations for the EoR (figure 1 in \citealt{Rahmati2018}). We interpolate to calculate the MFP for a derived neutral fraction from our simulation, and iterate to achieve a consistent ionization field and the MFP for each redshift step. As shown below, this model corresponds to a low level of the ionizing background and a short MFP.
\end{description}

\begin{figure}[!htbp]
\includegraphics[width=8.5cm,height=6.6cm]{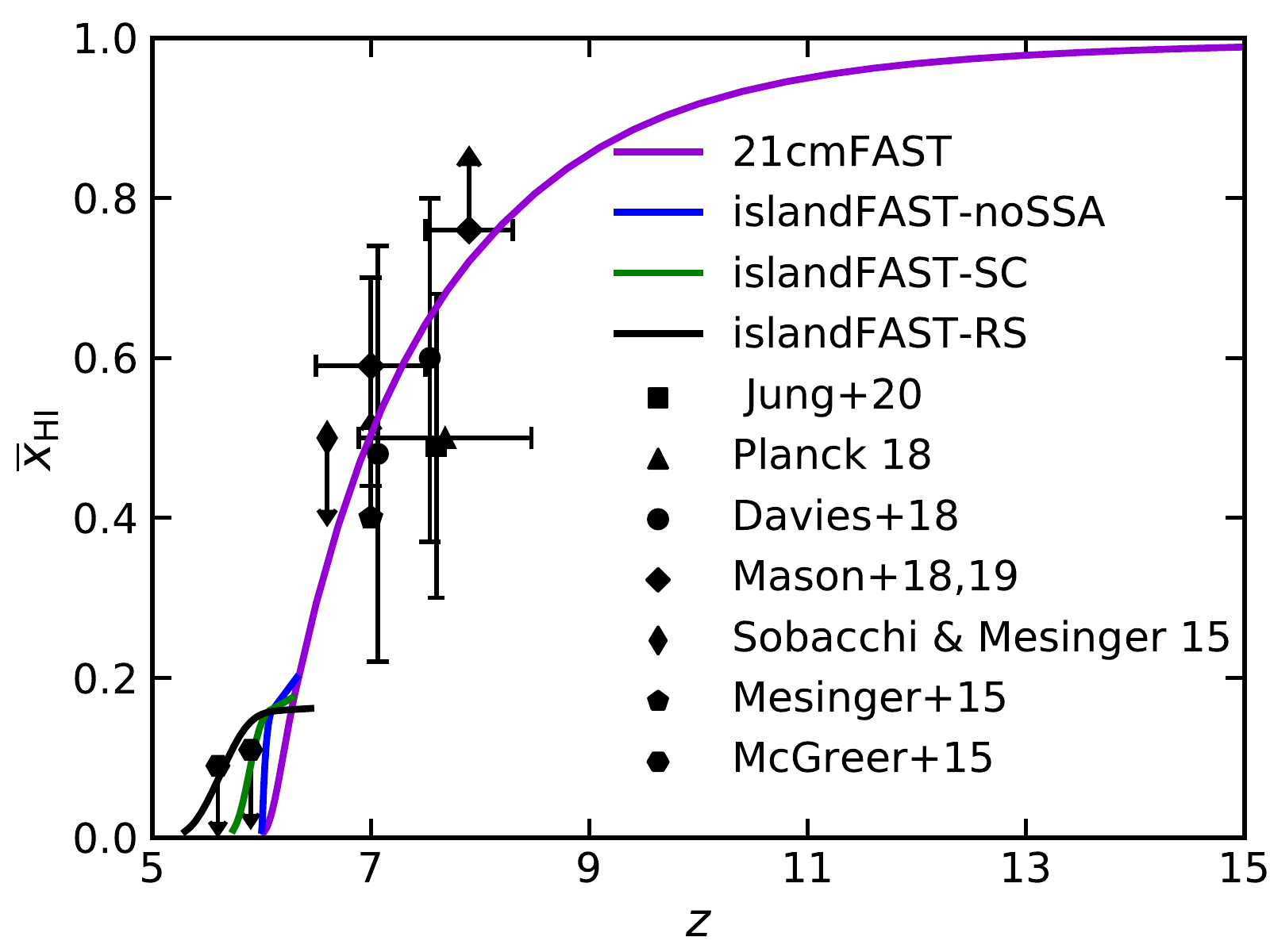}
\centering
\caption{The ionization history predicted by the four semi-numerical simulations (solid lines) as compared to various observation data. The observational data include constraints with the optical depth measurement from Planck \citep{Planck2020}, the neutral IGM damping wing on high-$z$ QSO spectra \citep{Davies2018}, the Ly$\alpha$ equivalent width distribution of Lyman-break galaxies \citep{Mason2018,Mason2019,2020ApJ...904..144J}, the clustering of Ly$\alpha$ emitters \citep{Sobacchi2015}, the Ly$\alpha$ fraction \citep{Mesinger2015}, and the dark pixel statistics of high-$z$ QSO spectra \citep{McGreer2015}, as indicated in the legend.}
\label{fig:history}
\end{figure}

\begin{figure*}[htbp]
\centering
\includegraphics[width=8cm]{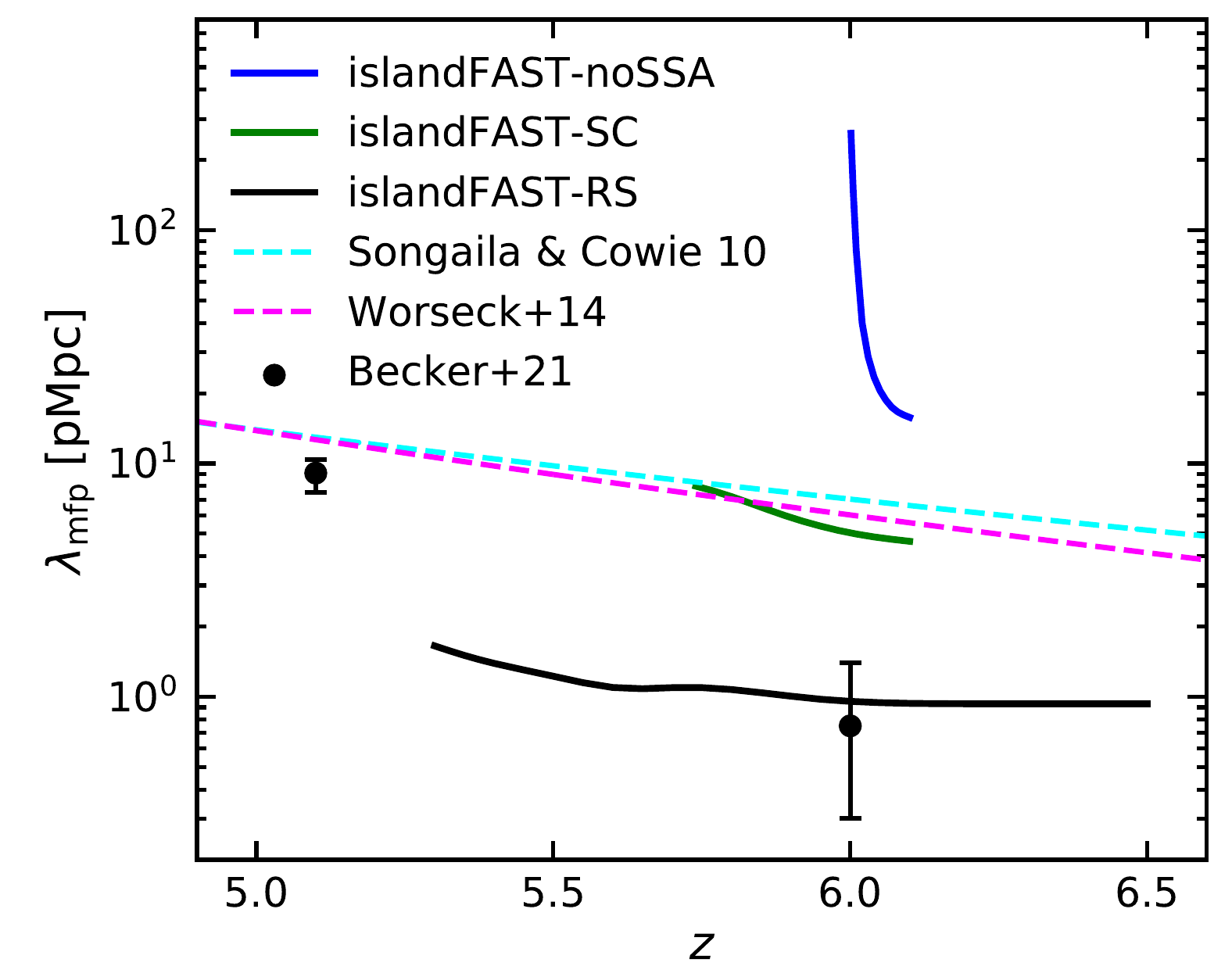}
\includegraphics[width=8cm]{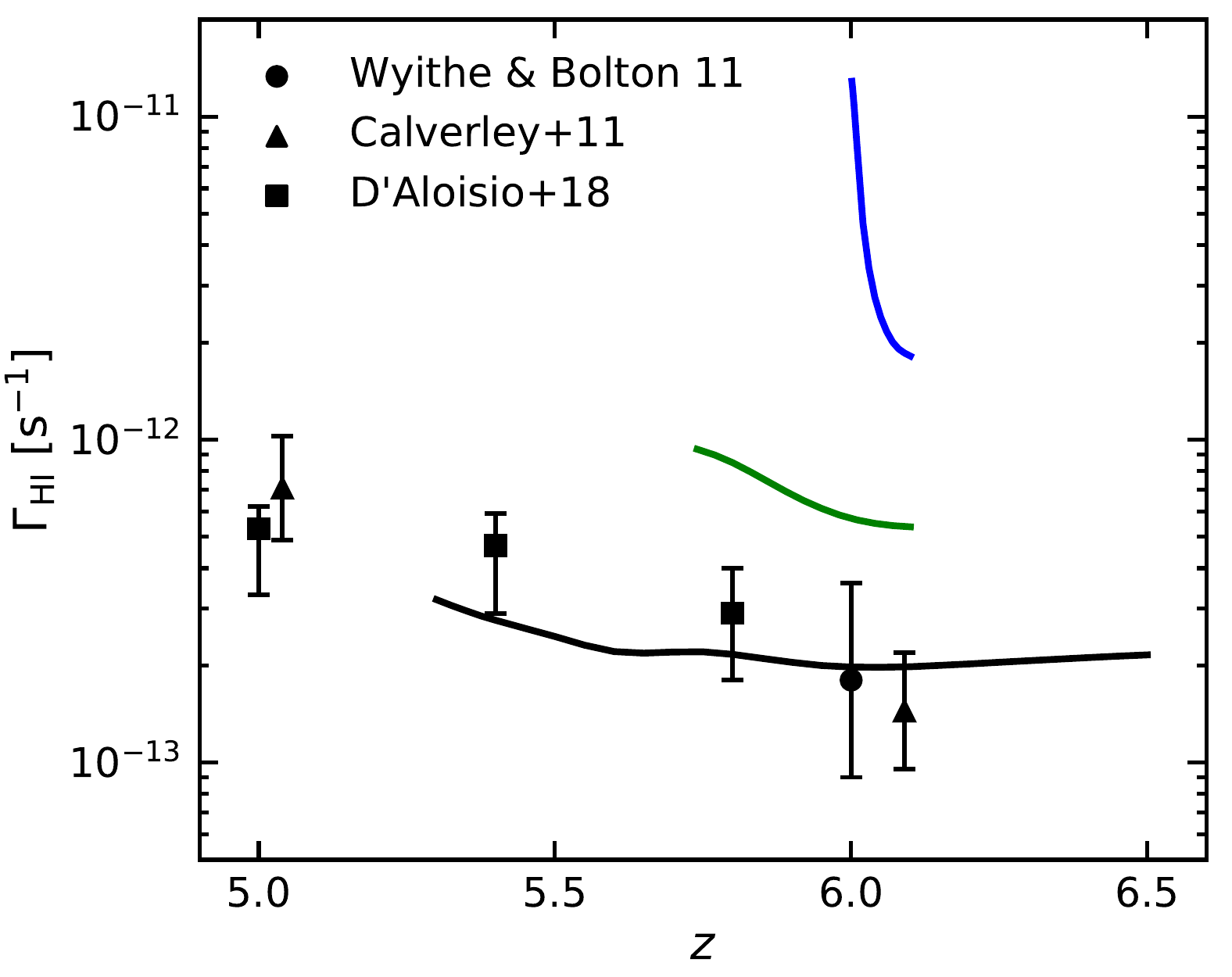}
\caption{{\it Left panel:} The evolution of the mean free path for ionizing photons. The blue, green and black solid lines refer to the evolutions from {\tt islandFAST-noSSA}, {\tt islandFAST-SC}, and {\tt islandFAST-RS} models, respectively.
The magenta dashed line shows the mean free path extrapolated from the fitting formula by \citet{Worseck2014}, while the cyan dashed line indicates the mean free path due to SSAs only, extrapolated from the fitting formula by \citet{Songaila2010}, and the dots with error bars show the recent MFP measurement by \cite{Becker2021}.
{\it Right panel:} The \textsc{H\,i} photoionization rate from the three simulations, compared with measurements from \citet{Wyithe2011}, \citet{Calverley2011}, and \citet{D'Aloisio2018}}
\label{fig:UVB}
\end{figure*}

Our simulation box has a comoving size of $1\, \rm Gpc$ a side and $600^3$ cells.
For each of the SSA model, we run two sets of simulations. One set adopts $\{\xi=10,\; T^{\rm vir}_{\rm min}=10^{4}\rm K\}$
corresponding to early and low-mass ionizing sources, and the other adopts $\{\xi=30,\; T^{\rm vir}_{\rm min}=5\times10^{4}\rm K\}$
corresponding to late and high-mass ionizing sources.
As a reference, we also run the {\tt 21cmFAST} for the same sets of parameters.
In the following analyses, the fiducial source parameters for {\tt 21cmFAST}, {\tt islandFAST-noSSA} and {\tt islandFAST-SC} models are
$\{\xi=10,\; T^{\rm vir}_{\rm min}=10^{4}\rm K\}$ , while the fiducial parameters for the {\tt islandFAST-RS} model are
$\{\xi=30,\; T^{\rm vir}_{\rm min}=5\times10^{4}\rm K\}$, chosen to be consistent with various observations.
The evolutions of the volume neutral fraction $\bar{x}_{\textsc{H\,i}}$ from the three {\tt islandFAST} models, and that from {\tt 21cmFAST},
are shown in Fig.~\ref{fig:history}. Also shown are the observational constraints from various observations as indicated in the legend and the caption.
The reionization processes predicted by the {\tt 21cmFAST}, {\tt islandFAST-noSSA}, {\tt islandFAST-SC} and {\tt islandFAST-RS} end at redshift $6.02$, $6.00$, $5.74$, and $5.30$, respectively. The corresponding Thompson optical depth are $0.0545$, $0.0543$, $0.0541$, and $0.0557$, respectively, all consistent with the latest Planck results \citep{Planck2020}.
We note that the reionization history of late EoR based on island-topology dose not connect smoothly with the history of early EoR simulated assuming bubble-topology. This is an intrinsic defect of the simulation in describing the complex percolation process.
However, as will be shown below (in Section \ref{size}), when compared at the same mean neutral fractions,
the {\tt islandFAST-noSSA} and {\tt 21cmFAST} predict consistent island statistics, so this discontinuity at the topological transition
would not affect the results on the island statistics.
We reserve the improvement on the simulation strategy for the mid-EoR to future works. Here we confirm the previous result that the presence of SSAs prolongs the reionization process.

Fig.~\ref{fig:UVB} shows the evolution of the MFP for ionizing photons (left panel) and the photoionization rate (right panel) predicted by the three SSA models. In the case of no SSA, the MFP is entirely limited by neutral islands, and as expected, the MFP grows rapidly as the islands being ionized (blue solid line), and the ionizing background also increases rapidly. In the presence of SSAs, the propagation of ionizing photons is effectively blocked, resulting in a significant reduction of the MFP and the ionizing background, and their growth rate also slows down (green solid line). The MFP gradually approaches the value completely limited by the SSAs (cyan dashed line). If we further enhance the absorption of SSAs to ionizing photons, as in the {\tt islandFAST-RS} model, the MFP and the ionizing background are further reduced, with only a weak evolution (black solid line). In this case, the reionization is significantly delayed by SSAs, and the island stage lasts for a long time.
The magenta dashed line is extrapolated from the direct measurement of the MFP using the drop in the continuum flux of high-resolution quasar spectra up to $z\sim 5$ \citep{Worseck2014}. It is seen that the {\tt islandFAST-SC} model is in the best agreement with the MFP measurements by \citet{Songaila2010}, \citet{Worseck2014}, and the data point at $z\sim5$ by \citet{Becker2021}, while the {\tt islandFAST-RS} model is more consistent with the photoionization rate measurements and the recent MFP measurement at $z\sim6$ by \citet{Becker2021}. Note that the observation data for the MFP and the ionizing background are still very limited at high redshifts, and the measurements could be biased due to the proximity effect \citep{DAloisio2018,Davies2020,Becker2021}. There is still large uncertainty in the MFP and the ionizing background beyond the reionization. The {\tt islandFAST-SC} model and the {\tt islandFAST-RS} model are both of great significance bracketing possible scenarios consistent with current observations.

\begin{figure*}[!htbp]
\includegraphics[width=12cm,height=12cm]{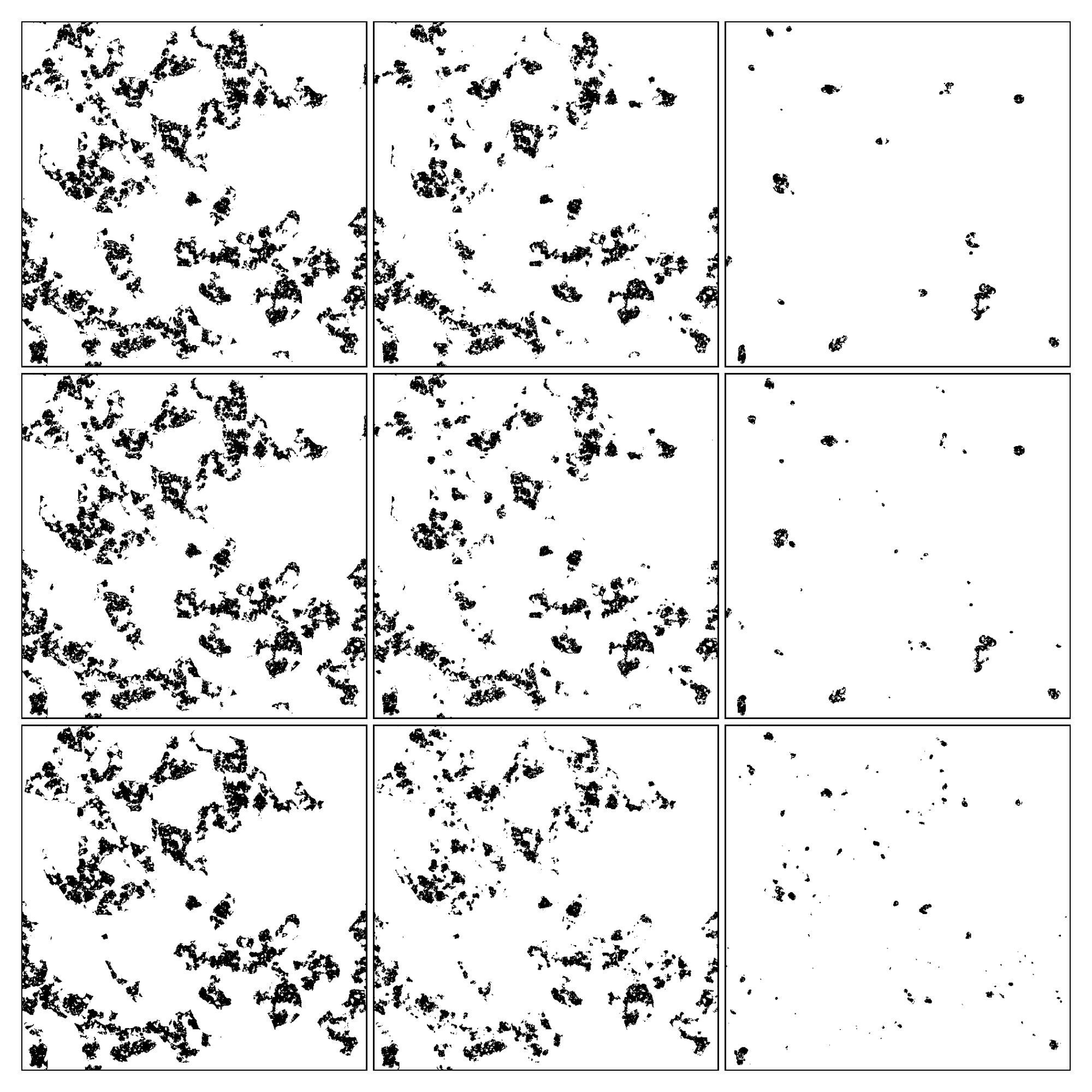}
\centering
\caption{Slices of the ionization fields from the {\tt islandFAST-noSSA}, {\tt islandFAST-SC}, and {\tt islandFAST-RS} simulations, top to bottom respectively. The neutral patches are shown as black, and the ionized regions are white. The three columns from left to right correspond to $\bar{x}_{\textsc{H\,i}}=0.16$, $0.10$, and $0.01$, respectively. All slices are $1\ \rm Gpc$ on a side and $1.67\ \rm Mpc$ thick in comoving scale.}
\label{fig:xHI-slices}
\end{figure*}

\section{Effects of absorption systems}\label{results}

Slices of the ionization field from the different semi-numerical simulations are shown in Fig.~\ref{fig:xHI-slices}. The top, middle and bottom panels are for {\tt islandFAST-noSSA}, {\tt islandFAST-SC}, and {\tt islandFAST-RS} models, respectively. All slices are 1 comoving Gpc on a side and 1.67 comoving Mpc thick, and the three columns correspond to mean neutral fractions of $0.16$, $0.10$, and $0.01$, from left to right respectively. The neutral islands are shown as black patches, and the ionized regions are left white. It is worth noting that most of the neutral islands are ``porous'', showing the ``bubbles-in-island'' effect. Generally, there is no significant morphological difference between the {\tt islandFAST-noSSA} model and the {\tt islandFAST-SC} model during late stage of reionization ($\bar{x}_{\textsc{H\,i}}\lesssim0.1$), and the remaining islands predicted by them are small in number but relatively large in scale. However, when the absorption of ionizing photons is significant enough, as in the {\tt islandFAST-RS} model, a larger number of small islands are left in the Universe. Different from the other two models, the topology of the neutral islands predicted by {\tt islandFAST-RS} model changes significantly between $\bar{x}_{\textsc{H\,i}}=0.16$ and $0.10$; the large islands break up into more small islands. It implies that, with the number density observation of SSAs up to higher redshifts in the future, we may be able to distinguish the reionization models with different morphology, according to the anti-correlation between the number density of SSAs and the typical size of neutral islands.

\begin{figure*}[p]     
\includegraphics[width=7cm]{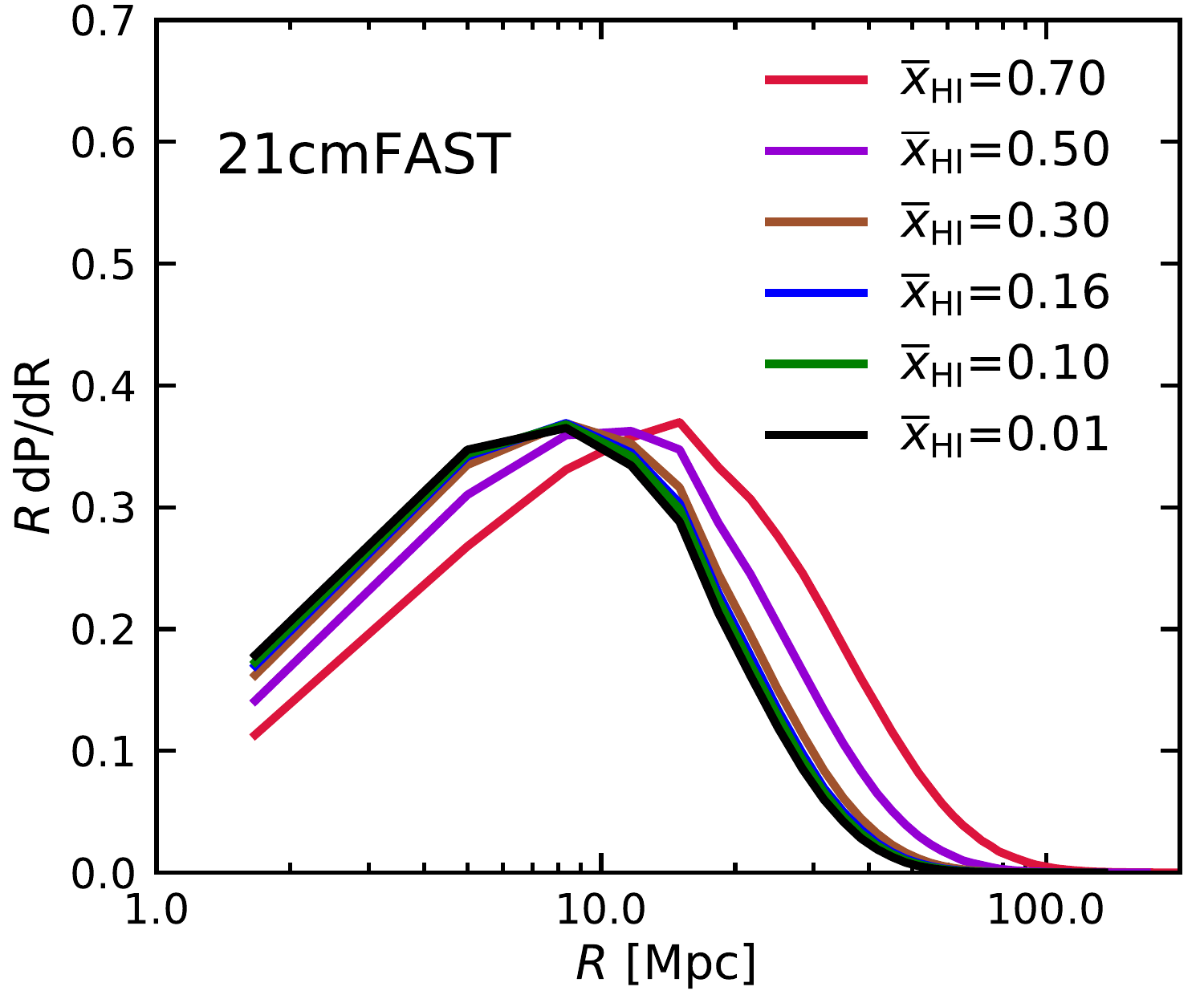}
\includegraphics[width=7cm]{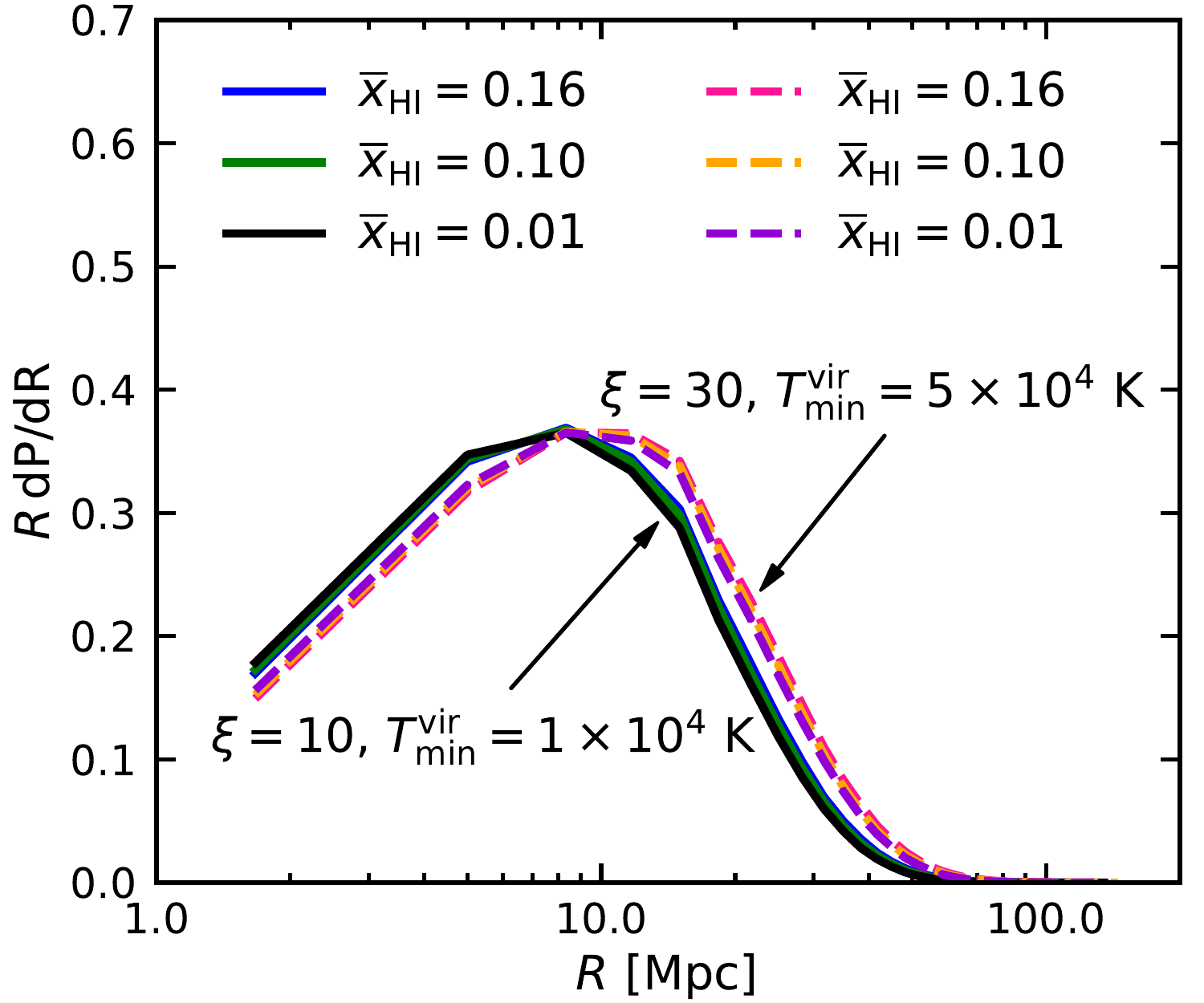}
\centering
\caption{Size distributions of neutral islands from {\tt 21cmFAST}. The curves in both panels are for different reionization stages with the mean neutral fractions indicated in the legend. The curves in the left panel from right to left show the results from early stage to late stage of reionization assuming $\{\xi=10,\, T^{\rm vir}_{\rm min}=10^{4}\rm K\}$. The right panel compares the results for the island stages between the simulation with $\{\xi=10,\, T^{\rm vir}_{\rm min}=10^{4}\rm K\}$ (solid lines) and the one with $\{\xi=30,\, T^{\rm vir}_{\rm min}=5\times10^{4}\rm K\}$ (dashed lines).}
\label{fig:size-21cmFAST}
\end{figure*}

\subsection{The size distribution of neutral islands}\label{size}

In this section, we explore the effects of absorption systems on the size distribution and its evolution of neutral islands. We characterize the scale of islands with the mean-free-path algorithm \citep{Mesinger2007}.  Note here the differences between the ``mean free path'' algorithm in characterizing the sizes of neural/ionized regions as the typical length of random vectors reaching a phase transition in the ionization field, and the MFP for ionizing photons to travel freely before being absorbed. As a reference model, we first present the size distributions of neutral regions from the {\tt 21cmFAST} simulation in Fig.~\ref{fig:size-21cmFAST}. Note that for the early stages of reionization, all the {\tt islandFAST} models are identical to the {\tt 21cmFAST}, before adopting the different topology for the island stage. For the island stage, the main difference between the original version of {\tt 21cmFAST} and the {\tt islandFAST-noSSA} model is the different assumption of the ionization topology and the different filtering algorithm, but both of them assumes no additional absorption systems.

\begin{figure*}[!htbp]
\centering
\includegraphics[width=5.7cm]{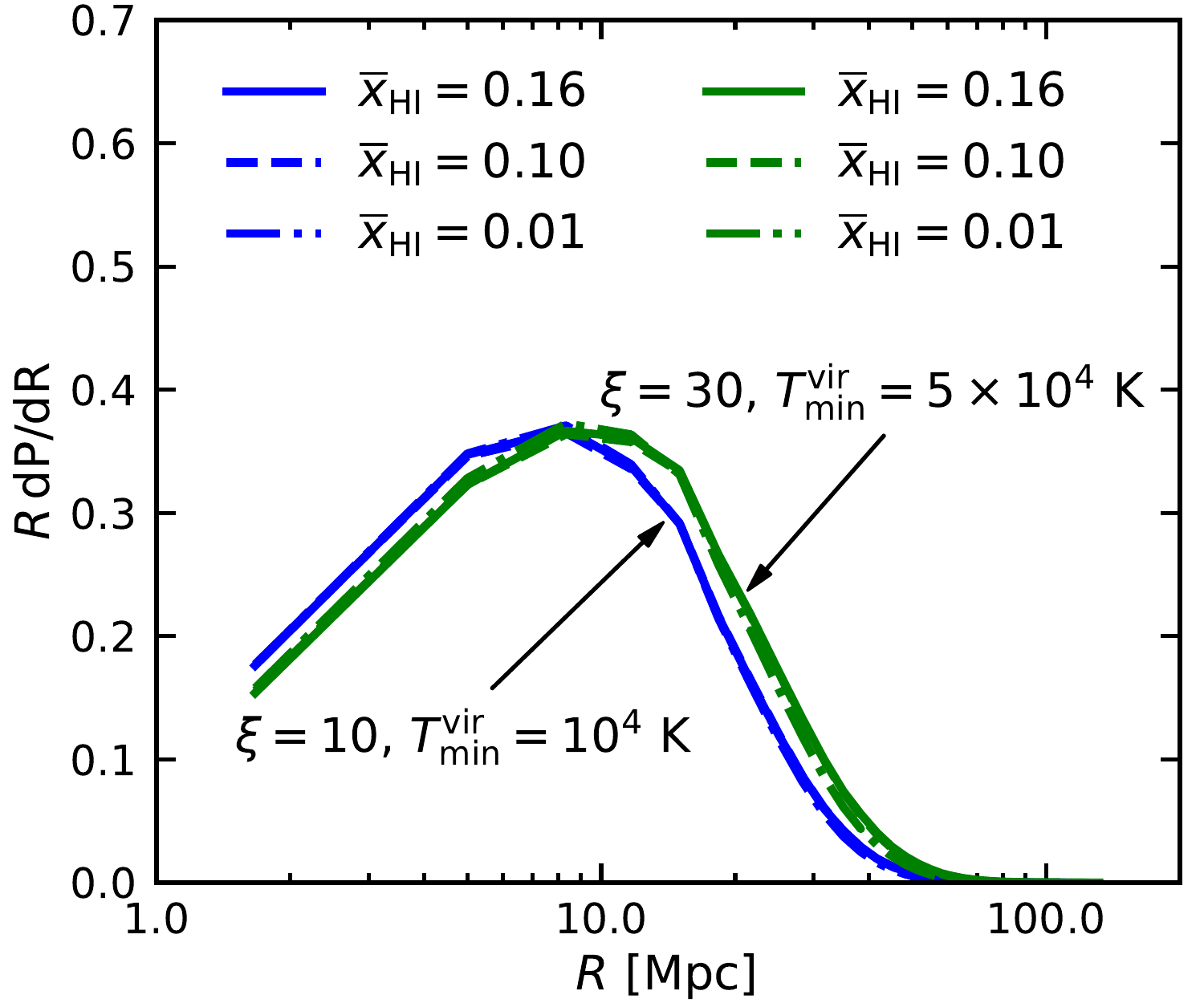}
\includegraphics[width=5.7cm]{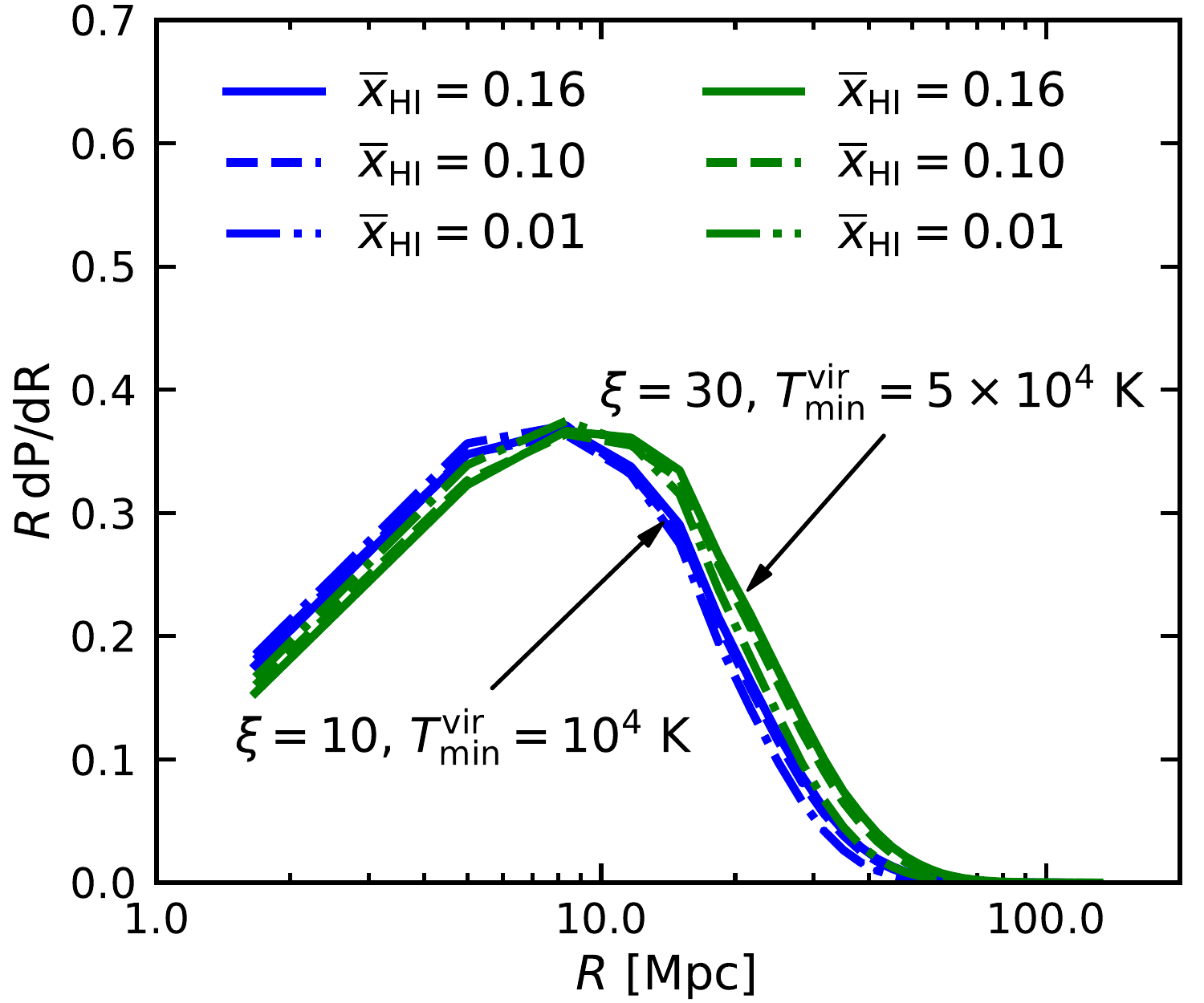}
\includegraphics[width=5.7cm]{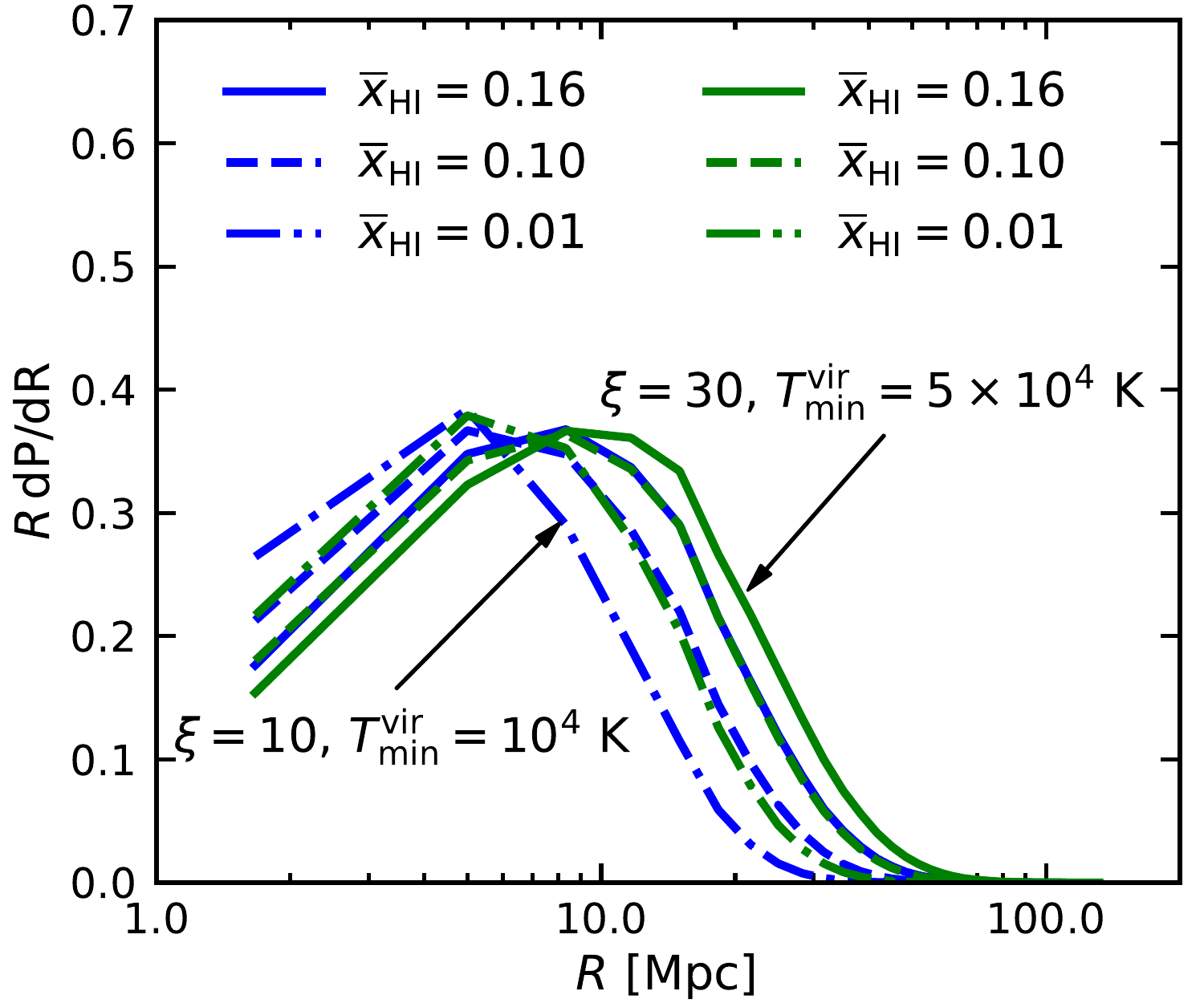}
\caption{Size distributions of neutral islands for the {\tt islandFAST-noSSA} model ({\it left panel}), {\tt islandFAST-SC} model ({\it middle panel}), and the {\tt islandFAST-RS} model ({\it right panel}).
The blue sets of curves correspond to source parameters of $\{\xi=10,\, T^{\rm vir}_{\rm min}=10^{4}\rm K\}$, and the green curves are for $\{\xi=30,\, T^{\rm vir}_{\rm min}=5\times10^{4}\rm K\}$.
The solid, dashed, and dot-dashed lines are for $\bar{x}_{\textsc{H\,i}}=0.16$, $\bar{x}_{\textsc{H\,i}}=0.10$, and $\bar{x}_{\textsc{H\,i}}=0.01$, respectively.}
\label{fig:size-models}
\end{figure*}

\begin{figure*}[!htbp]
\centering
\includegraphics[width=5.7cm]{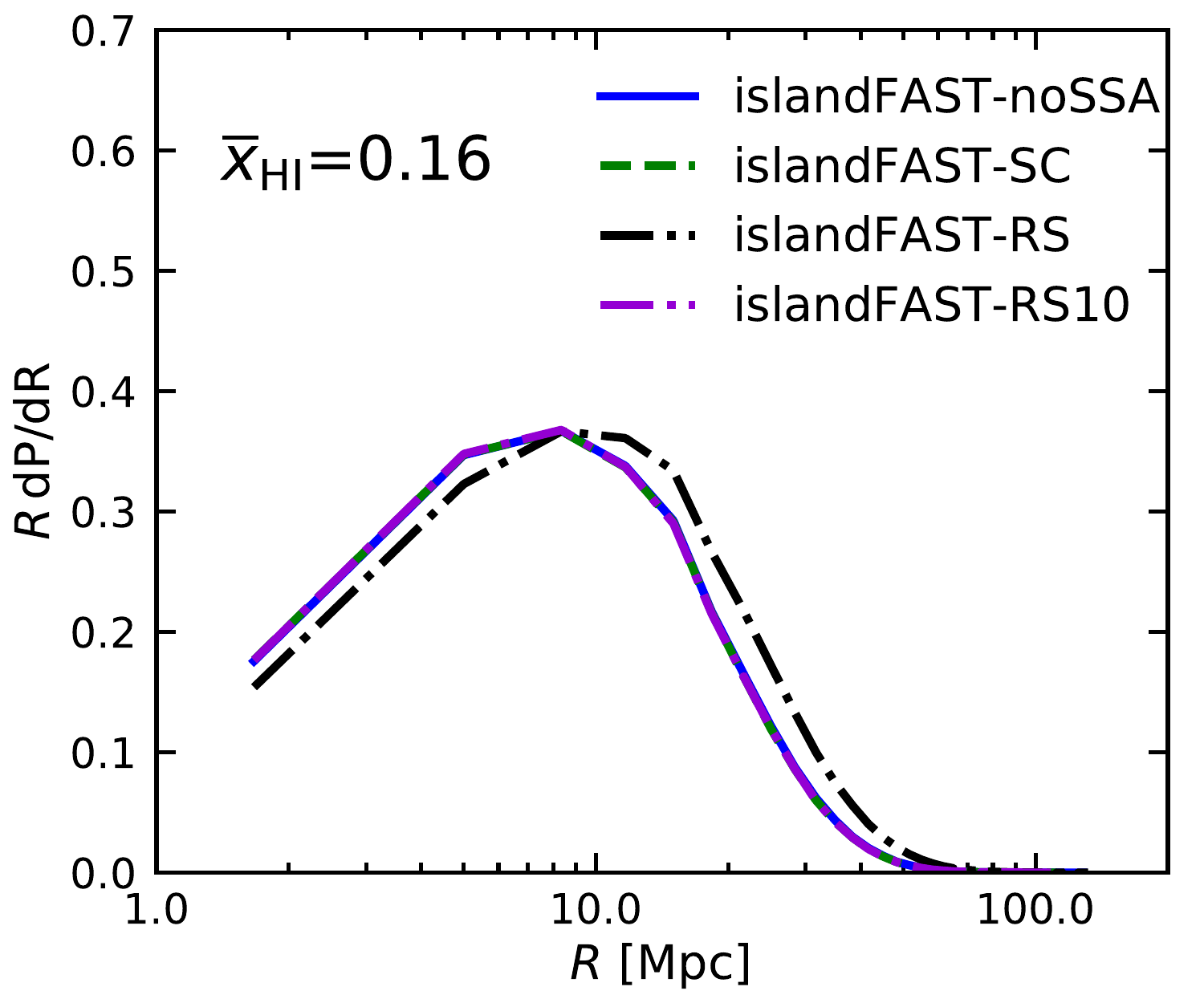}
\includegraphics[width=5.7cm]{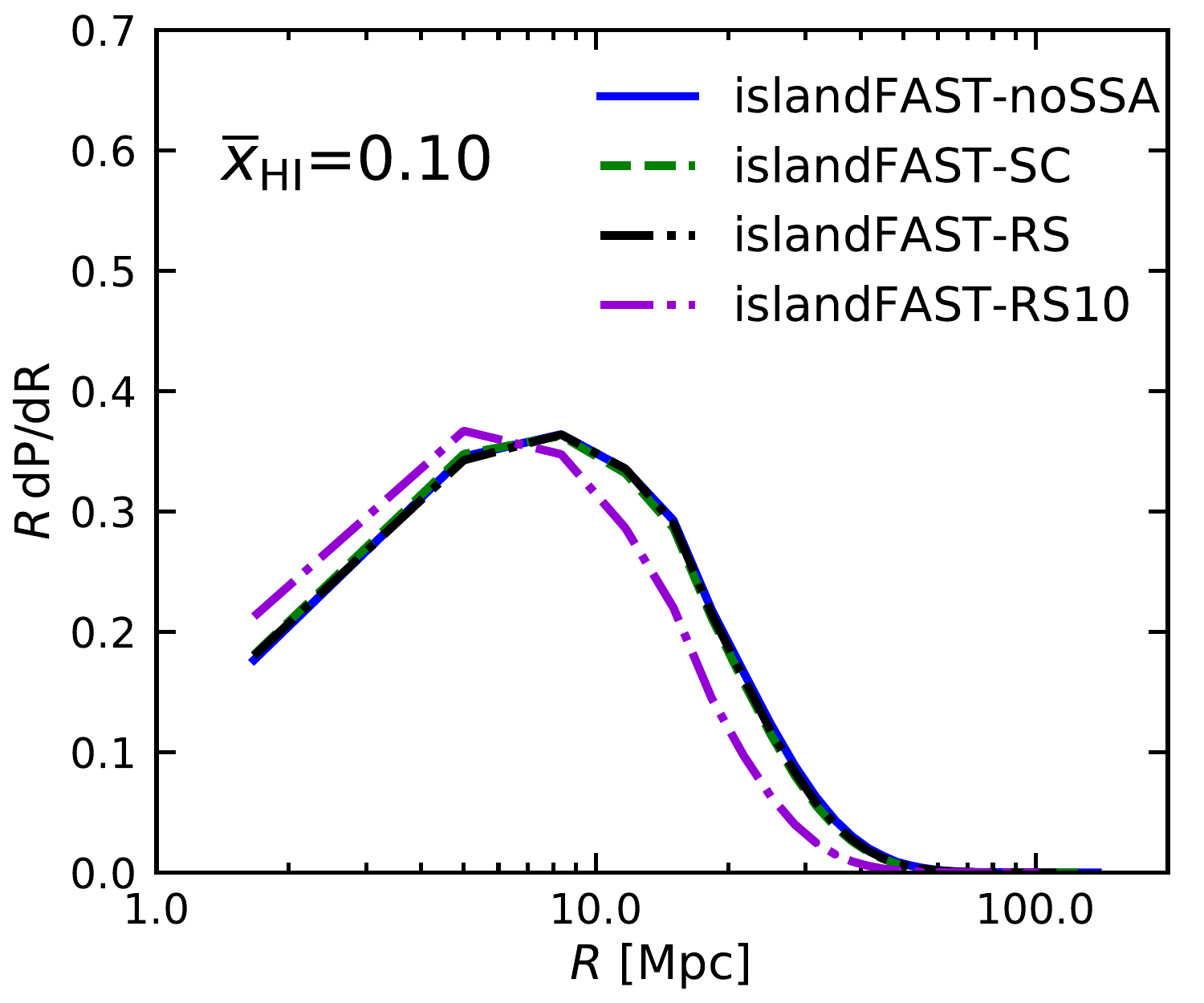}
\includegraphics[width=5.7cm]{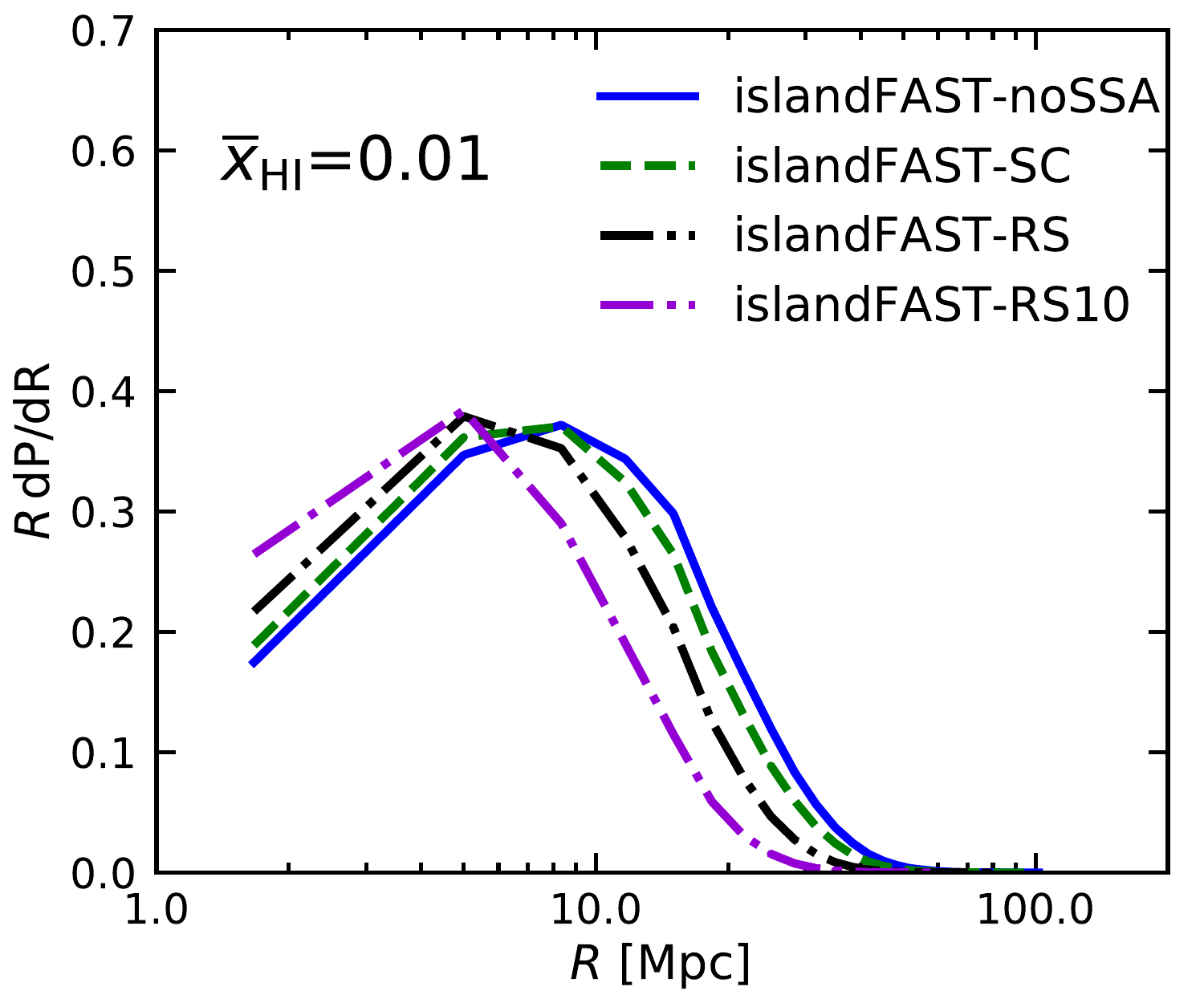}
\caption{Size distributions of neutral islands at $\bar{x}_{\textsc{H\,i}}=0.16$ ({\it left panel}), $\bar{x}_{\textsc{H\,i}}=0.10$ ({\it middle panel}), and $\bar{x}_{\textsc{H\,i}}=0.01$ ({\it right panel}). The blue solid, green dashed, and purple dot-dashed curves are for the {\tt islandFAST-noSSA}, {\tt islandFAST-SC}, and {\tt islandFAST-RS} models, respectively, with source parameters of $\{\xi=10,\, T^{\rm vir}_{\rm min}=10^{4}\rm K\}$, while the black dot-dashed curves are for {\tt islandFAST-RS} model with $\{\xi=30,\, T^{\rm vir}_{\rm min}=5\times10^{4}\rm K\}$.}
\label{fig:size-nf}
\end{figure*}

The left panel of Fig.~\ref{fig:size-21cmFAST} shows the evolution of the size distribution adopting the parameters of $\{\xi=10,\; T^{\rm vir}_{\rm min}=10^{4}\rm K\}$. The size distribution of neutral regions evolves during the early and mid stages of EoR, however, when the Universe enters the neutral fibers stage ($\bar{x}_{\textsc{H\,i}}\lesssim0.3$) \citep{ChenMF2019}, the evolution becomes extremely slow and the characteristic scale stalls at about $10$ comoving Mpc. This is consistent with the intuition that when measuring the size of neutral regions with the mean-free-path algorithm, most random vectors would not be parallel with the stretching direction of neutral fibers, and the derived size would be similar to the neutral islands that the neutral fibers fragment into later. However, the size distribution will show a more significant evolution from the neutral fiber stage to the island stage if one characterizes the size using the Friend-of-Friend algorithm, as illustrated in \citet{Giri2019}.

The right panel compares the results between the simulation with $\{\xi=10,\; T^{\rm vir}_{\rm min}=10^{4}\rm K\}$, and the one with $\{\xi=30,\; T^{\rm vir}_{\rm min}=5\times10^{4}\rm K\}$ during the late EoR. We find that the characteristic scale of neutral regions does depend on the properties of ionizing sources. More massive ionizing sources with higher ionizing efficiencies will lead to larger islands during the late stage, which is consistent with \citet{Giri2019}.

In Fig.~\ref{fig:size-models}, we present the size distributions of the neutral islands for the three SSA models, during the island stage of reionization.
In each panel, the blue and green sets of curves correspond to source parameters
of $\{\xi = 10,\; T^{\rm vir}_{\rm min}=10^{4}\rm K\}$ and $\{\xi=30,\; T^{\rm vir}_{\rm min}=5\times10^{4}\, \rm K\}$ respectively.
Though with the even faster reionization process, the evolution and source dependence of the island size distribution from the
{\tt islandFAST-noSSA} model (left panel of Fig.~\ref{fig:size-models}) are similar to the behavior predicted by {\tt 21cmFAST} as shown in Fig. \ref{fig:size-21cmFAST}.
We note that the source dependence of the island size distribution is similar among the different SSA models,
with the more massive ionizing sources resulting in larger typical size of islands.
However, the different SSA models predict distinct evolutionary behavior of the island size distribution.
 We find that the characteristic island scale is almost unchanged and stalls at $\sim 10$ comoving Mpc in the {\tt islandFAST-noSSA} model and the {\tt islandFAST-SC} model, which have a strong ionizing background or a long MFP, but it shows a clear evolution in the {\tt islandFAST-RS} model with weak ionizing background or a short MFP.
This reflects the competitive roles played by the ionizing background and the ionizing sources inside the islands. When the ionizing background dominates the ionizing budget, the islands will be rapidly ionized from the outside, and it would be hard for the small islands to keep their identities. However, if the inside ionizing sources dominates the ionizing photons, the islands tend to disintegrate from the inside, which results in the more obvious evolution in the island size and a more prolonged reionization process because of the weaker ionizing background.

We compare the size distributions of neutral islands between the three models at the same neutral fractions in Fig.~\ref{fig:size-nf}. The blue solid, green dashed, and purple dot-dashed curves are for the {\tt islandFAST-noSSA}, {\tt islandFAST-SC}, and {\tt islandFAST-RS} models, respectively, with source parameters of $\{\xi=10, \,T^{\rm vir}_{\rm min}=10^{4}\rm K\}$, 
while the black dot-dashed curves are for {\tt islandFAST-RS} model with $\{\xi=30, \,T^{\rm vir}_{\rm min}=5\times10^{4}\rm K\}$.
With the same source parameters, all the models predict the same island size distribution at $\bar{x}_{\textsc{H\,i}}=0.16$.
However, in the {\tt islandFAST-RS} model, the size distribution of islands moves to smaller scales with the decreasing neutral fraction, and at the very late stage ($\bar{x}_{\textsc{H\,i}}\sim 0.01$), the characteristic island scale predicted by the {\tt islandFAST-RS} model is obviously smaller.
Coincidently, the size distribution in the {\tt islandFAST-RS} model with
$\{\xi=30$, $T^{\rm vir}_{\rm min}=5\times10^{4}\, \rm K\}$ is almost identical to that predicted by the other two models with
$\{\xi=10$, $T^{\rm vir}_{\rm min}=10^{4}\, \rm K\}$ at $\bar{x}_{\textsc{H\,i}}=0.10$, showing the degeneracy between the
source properties and the absorber abundances. Although the effect of lower-mass sources with lower ionizing efficiency could
mimic the effect of having more abundant SSAs at a certain redshift, this degeneracy can be easily broken
by observing multiple redshifts and making use of the evolutionary behavior of the island size distribution.

\begin{figure}[t]   
\includegraphics[scale=0.45]{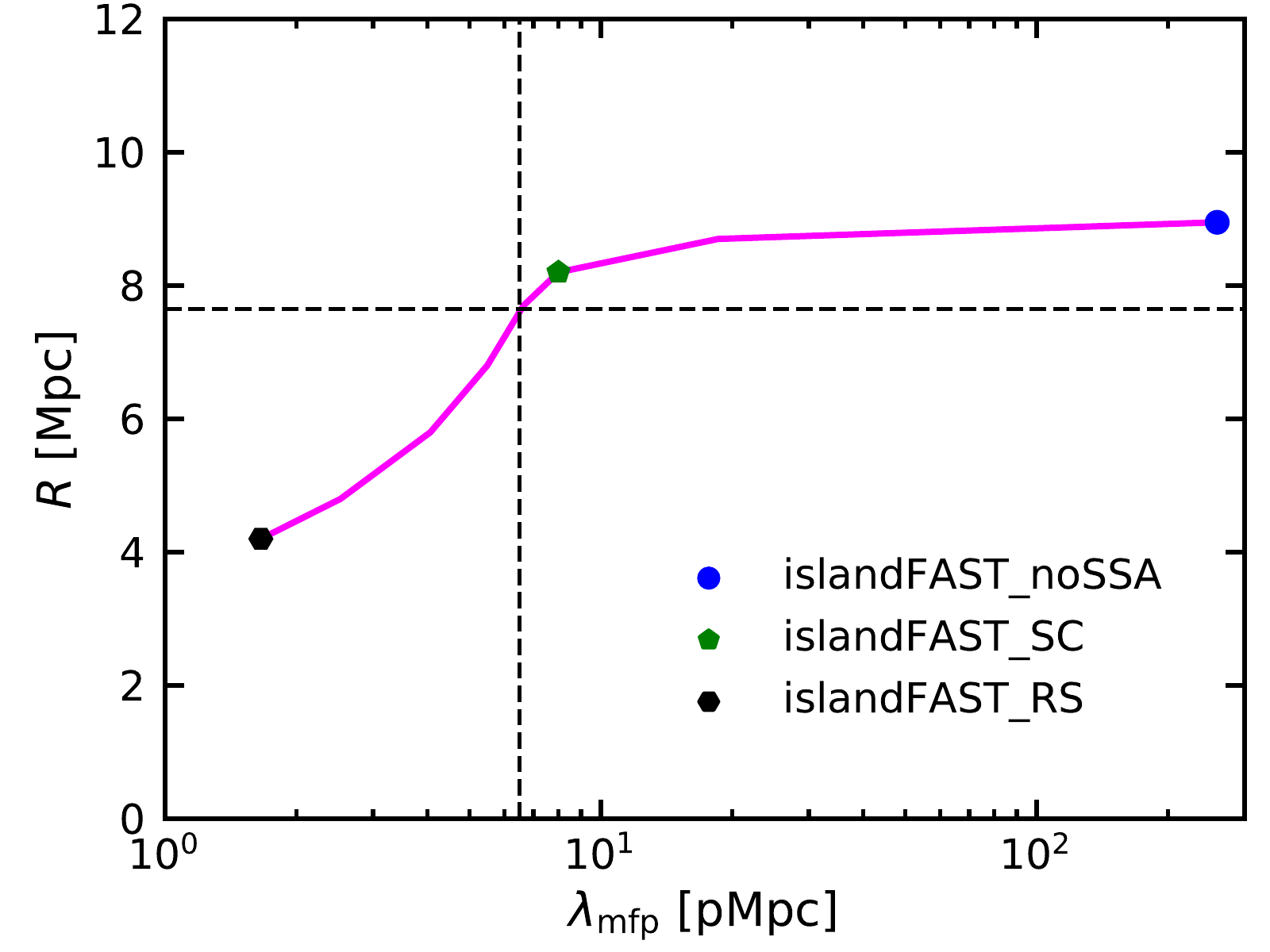}
\centering
\caption{The relation between the characteristic island scale and $\lambda_{\rm mfp}$ at the ending stage of reionization when $\bar{x}_{\textsc{H\,i}}=0.01$. The {\tt islandFAST-noSSA}, {\tt islandFAST-SC} and {\tt islandFAST-RS} models are shown with blue dot, green pentagon and black hexagon, respectively. The horizontal dashed line represents a scale $15\%$ smaller than the characteristic island scale predicted by {\tt islandFAST-noSSA}, and the vertical dashed line denotes the corresponding $\lambda_{\rm mfp}$.}
\label{fig:R_mfp}
\end{figure}

We notice that there is no significant difference in the size distribution between the {\tt islandFAST-noSSA} model and the {\tt islandFAST-SC} model, 
although the level of the ionizing background differs significantly as shown in Fig.~\ref{fig:UVB}, 
indicating that the characteristic island scale is not sensitive to the abundance of the SSAs, at least below a certain abundance threshold.
In order to estimate this threshold, we perform several {\tt islandFAST-SC} simulations with the fiducial parameters of $\{\xi=10,\; T^{\rm vir}_{\rm min}=10^{4}\rm K\}$,
varying the prefactor in Eq.~(\ref{eq.mfp_SC}); $\lambda_{\rm abs}$ is taken as 3, 0.8, 0.6, 0.4 and 0.2 times of the fiducial {\tt islandFAST-SC} model, respectively.
Then we measure the characteristic island scale, at which the size distribution peaks, at $\bar{x}_{\textsc{H\,i}}=0.01$ for each simulation.
Fig.~\ref{fig:R_mfp} shows the relation between the characteristic island scale and $\lambda_{\rm mfp}$. 
It is seen that as the abundance of SSAs decreases, or the MFP increases, the characteristic island size first increases,
and then gradually saturates at about 9 comoving Mpc when $\lambda_{\rm mfp} \gtrsim 10$ physical Mpc.
The SSA abundance in the fiducial {\tt islandFAST-SC} model roughly represents the abundance threshold in this case.
The horizontal dashed line indicates the scale $15\%$ smaller than that predicted by the {\tt islandFAST-noSSA} model,
and the corresponding $\lambda_{\rm mfp}$ is about $6.5$ physical Mpc (vertical dashed line). 
Note that at the very late stage ($\bar{x}_{\textsc{H\,i}}\lesssim0.01$), the SSAs dominate the opacity of the IGM, 
and $\lambda_{\rm abs}$ can be approximated as $\lambda_{\rm mfp}$, 
then the SSA abundance can be effectively constrained by measuring the typical scale of the remaining neutral islands.

\subsection{Long Troughs}

\cite{Becker2015} observed an extremely long ($\sim150$ comoving Mpc) opaque Gunn-Peterson trough extending down to redshift $5.5$ in the spectrum of quasar ULAS J0148+0600. Several works have proposed that there could be large islands persist till the late EoR, giving rise to a non-negligible possibility of lines of sight passing through these large islands which results in the long troughs as observed \citep{Kulkarni2019,Giri2019,Keating2020}.
Here we also calculate the probability that the observed long absorption trough is caused by the neutral islands in the late EoR.

\begin{figure}[tbp]
\includegraphics[scale=0.45]{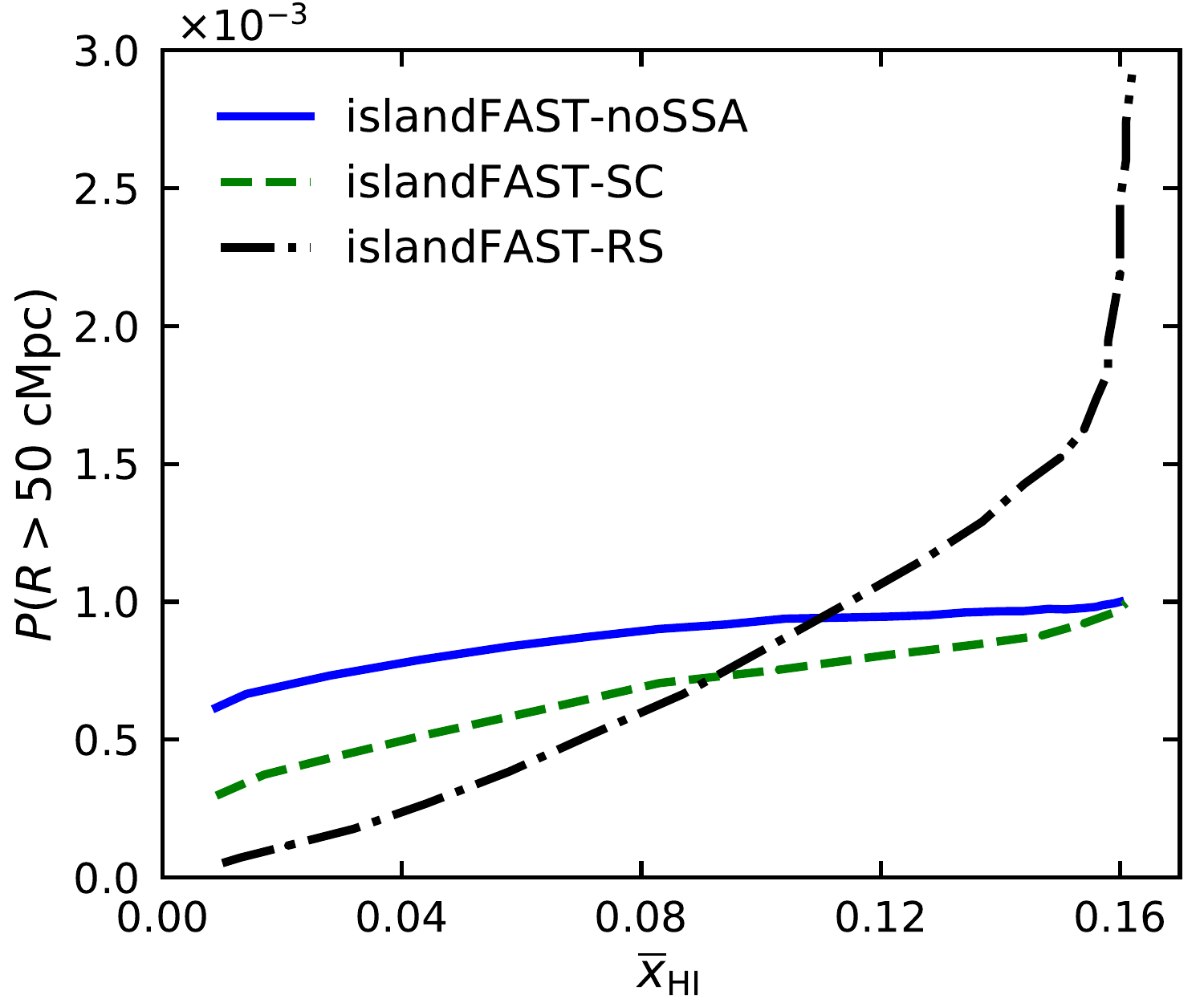}
\centering
\caption{The probability of the length of a line of sight intersected by a neutral island being larger than $50$ comoving Mpc. The solid, dashed, and dot-dashed lines are for the results from {\tt islandFAST-noSSA}, {\tt islandFAST-SC}, and {\tt islandFAST-RS} simulations, respectively. }
\label{fig:max}
\end{figure}

Fig.~\ref{fig:max} shows the probability for the length of a sightline intersected by an island to be larger than $50$ comoving Mpc,
as a function of the mean neutral fraction, from the simulated three models.
This represents the lower limit to the islands contribution to the Ly-$\alpha$ absorption trough, where the neutral islands
are the only contributor to the long troughs.
We find that for all the three models, the possibility of having an island scale larger than $50$ comoving Mpc is quite low (of order $\sim 10^{-3}$) for the island stage, but the probability increases rapidly towards the neutral fiber and earlier stages. This implies that if the long trough is solely caused by a large neutral region, the reionization has to complete quite late.

\begin{figure}[t]
\includegraphics[scale=0.45]{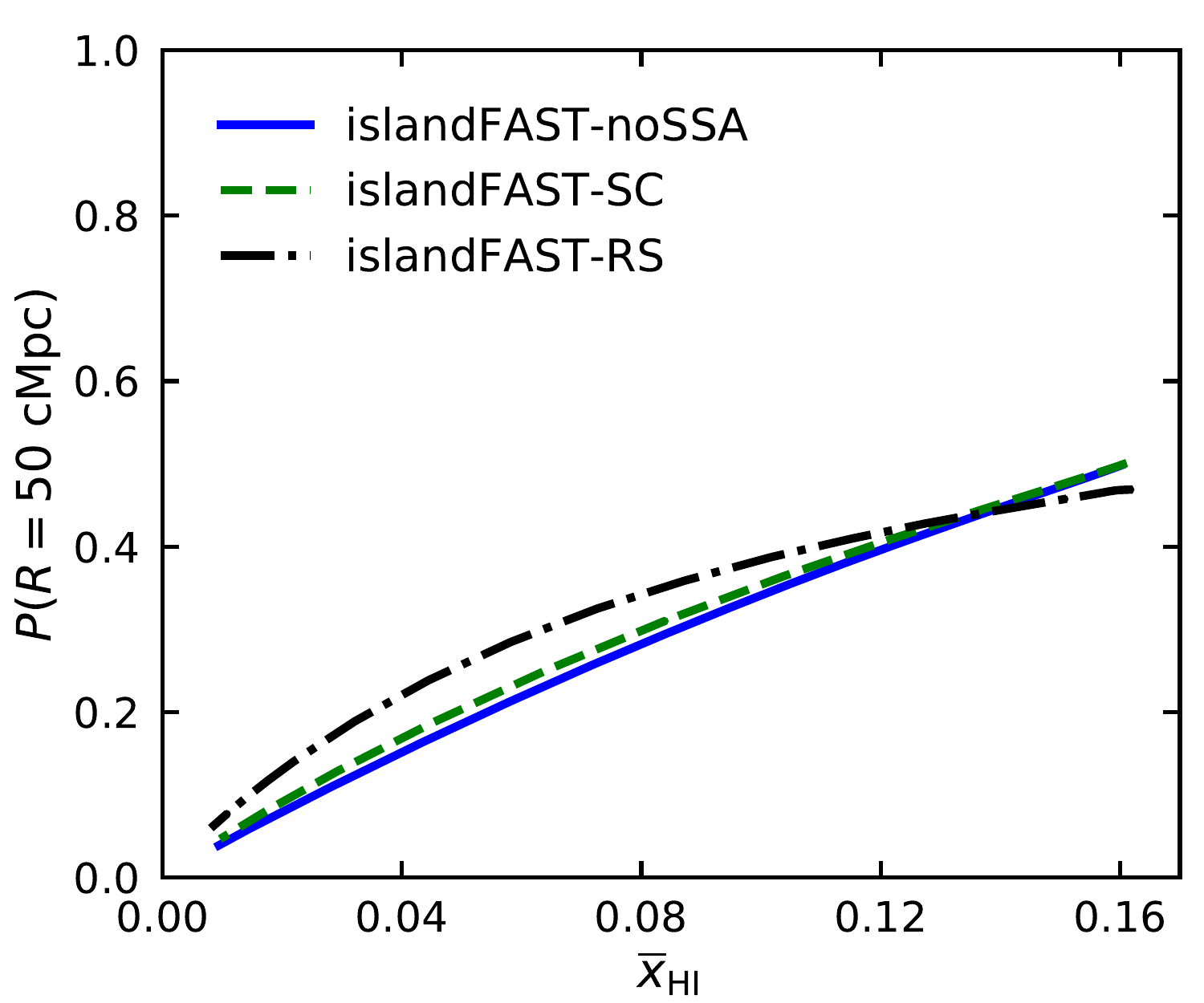}
\centering
\caption{The probability of a random $50$ comoving Mpc sightline intersecting a neutral island. The solid, dashed, and dot-dashed lines are for the results from {\tt islandFAST-noSSA}, {\tt islandFAST-SC}, and {\tt islandFAST-RS} simulations, respectively.}
\label{fig:p50}
\end{figure}

However, our model does not include partial ionizations, and the bubbles-in-islands significantly reduce the length scale of the intersected lines of sight passing through the porous neutral islands. The residual neutral hydrogen in the bubbles-in-islands may connect the neighboring absorption troughs giving rise to longer troughs.
In addition, the damping wing absorption of neutral islands can produce troughs longer than the islands themselves.
On the other hand, the residual neutral hydrogen in the ionized regions could also contribute to the Ly-$\alpha$ opacity, and may significantly enlarge the absorption troughs. The opacity of the ionized IGM is closely related to the residual neutral fraction of gas at low density \citep{Oh2005}. The ionization equilibrium at a fixed gas density results in a local neutral fraction depending on the photoionization rate of hydrogen $\Gamma_{\textsc{H\,i}}$ and the gas temperature $T$, i.e. $x_{\textsc{H\,i}}\propto {\Gamma}^{-1}_{\textsc{H\,i}}T^{-0.7}$ \citep{Oh2005}.
Therefore, a lower temperature or a lower photoionization rate will lead to the increase of the residual neutral fraction. The large-scale over-dense regions are likely to be cooler than average, because they were reionized early \citep{D'Aloisio2015}, and large-scale voids are likely to have a weaker ionizing background \citep{Davies2016}. These regions could both be mostly ionized, but still have a residual neutral fraction high enough to saturate the absorption of Ly$\alpha$ photons \citep{Keating2020}. We plan to investigate these possibilities by incorporating fluctuations in the ionizing background in future works.
Here we calculate the probability of a random 50 comoving Mpc sightline intersecting an island of any size, and the results are shown in Fig.~\ref{fig:p50}. These curves put upper limits on the contribution of islands to such troughs in different SSA models. We see that {\tt islandFAST-RS} provides the lowest probability at $\bar{x}_{\textsc{H\,i}}=0.16$ due to the higher fiducial parameters of
$\{\xi=30$, $T^{\rm vir}_{\rm min}=5\times10^{4}\, \rm K\}$, which result in larger and fewer islands, but thereafter, as the large islands break into many small ones, it provides the highest upper limits of the probability.

\begin{figure*}[htbp]
\includegraphics[width=5.7cm,height=4.9cm]{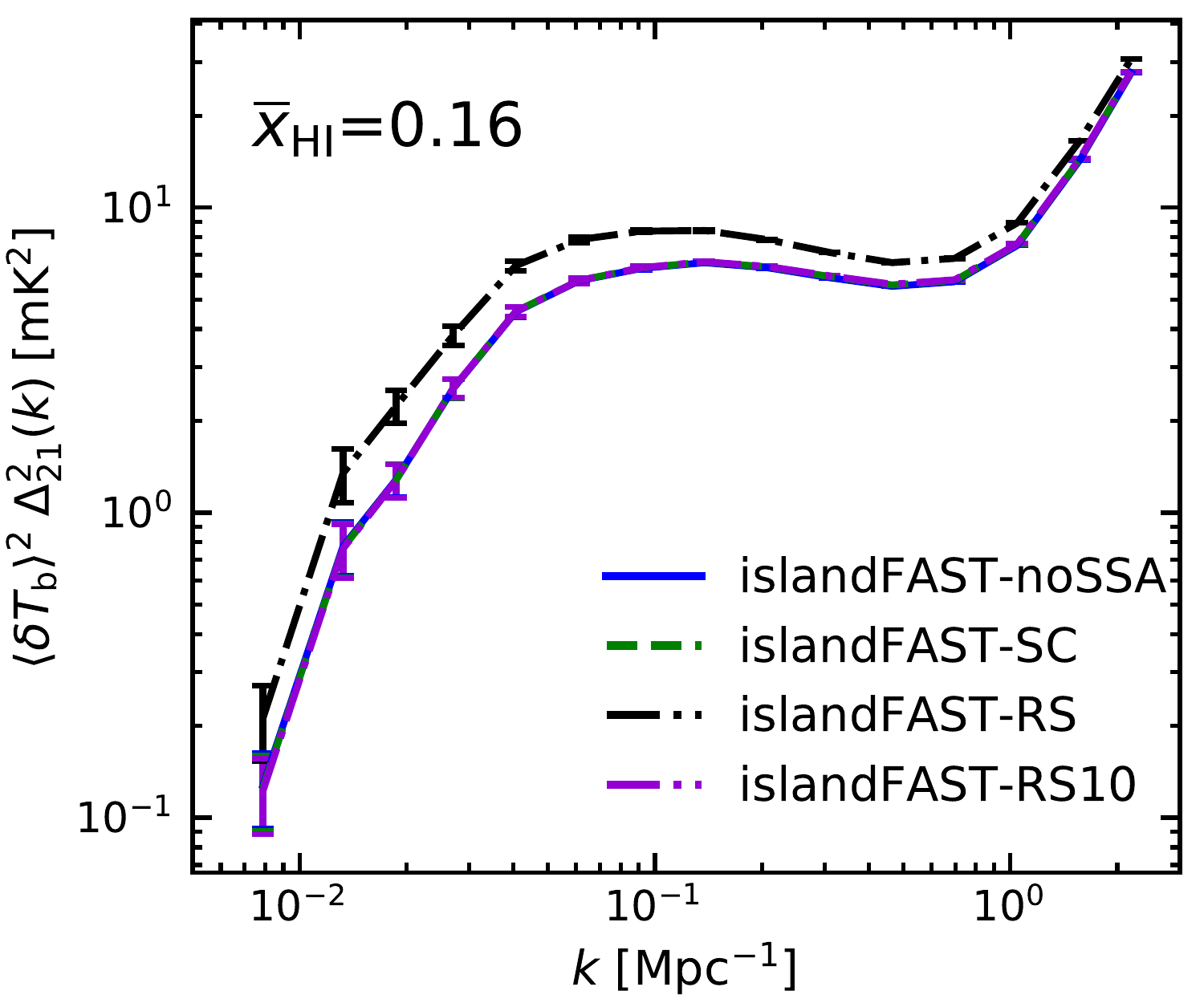}
\includegraphics[width=5.7cm,height=4.9cm]{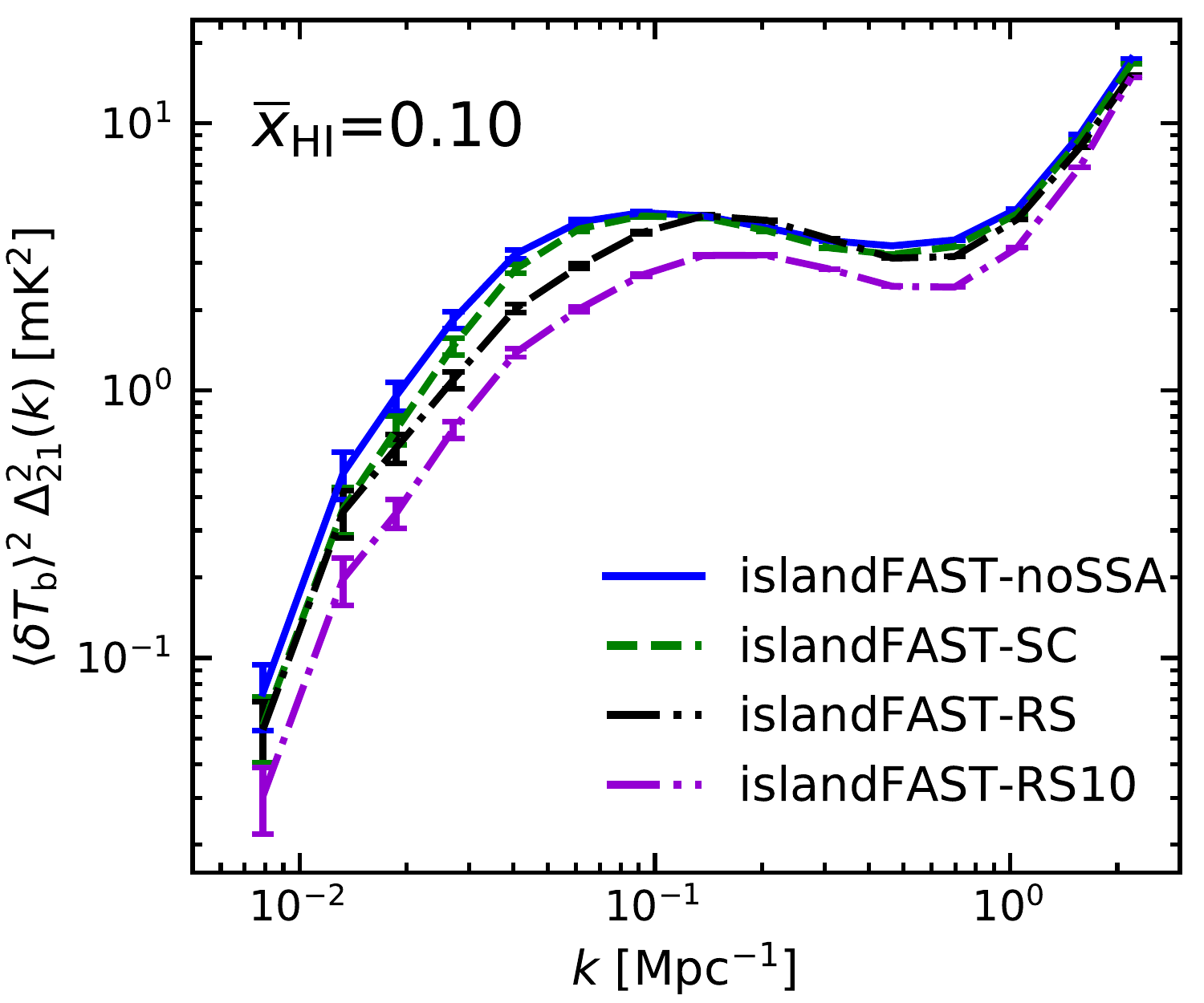}
\includegraphics[width=5.7cm,height=4.9cm]{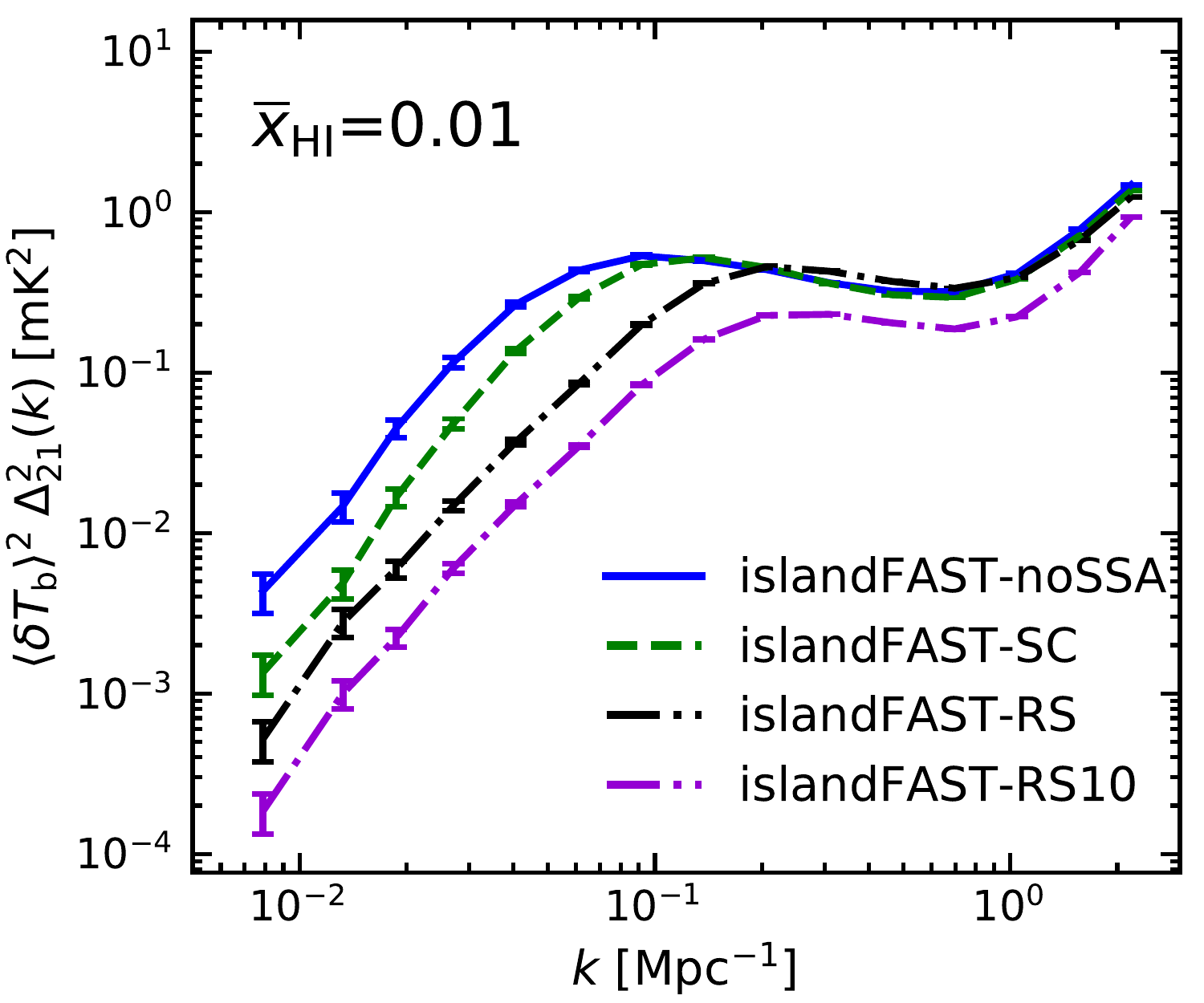}
\centering
\caption{The 21-cm power spectra with statistical errors at $\bar{x}_{\textsc{H\,i}}=0.16$ ({\it left panel}), $\bar{x}_{\textsc{H\,i}}=0.10$ ({\it middle panel}), and $\bar{x}_{\textsc{H\,i}}=0.01$ ({\it right panel}).
The blue solid, green dashed, and purple dot-dashed curves are for the {\tt islandFAST-noSSA}, {\tt islandFAST-SC}, and {\tt islandFAST-RS} models, respectively, with source parameters of $\{\xi=10,\ T^{\rm vir}_{\rm min}=10^{4}\rm K\}$, while the black dot-dashed curves are for {\tt islandFAST-RS} model with $\{\xi=30,\ T^{\rm vir}_{\rm min}=5\times10^{4}\rm K\}$.}
\label{fig:power-spectrum}
\end{figure*}

\subsection{The 21-cm Power Spectrum}\label{21cmps}
The effect of SSAs on the morphology of neutral islands may also be reflected in the 21-cm power spectrum, hopefully to be constrained in the near future. The differential 21-cm brightness temperature of hydrogen against the CMB at redshift $z$, corresponding to the observed frequency $\nu$, can be written as (e.g. \citealt{21CMFAST2011})
\begin{align}
\delta T_{\rm b}(\nu)&=27x_{\textsc{H\,i}}(1+{\delta}_{\rm nl})\bigg(\frac{H}{dv_{r}/dr + H}\bigg)\bigg(1-\frac{T_{\gamma}}{T_{\rm S}}\bigg)\n\\
&\quad\ \times \bigg( \frac{1+z}{10} \frac{0.15}{\Omega_{\rm m}h^2} \bigg)^{1/2} \bigg( \frac{\Omega_{\rm b}h^2}{0.023} \bigg) \rm mK,
\end{align}
where $x_{\textsc{H\,i}}$ is the neutral fraction, $\delta_{\rm nl}(\mathbf{x}, z) \equiv \rho / \bar{\rho} - 1$ is the evolved density contrast, 
$dv_{r}/dr$ is the comoving gradient of the line of sight component of the comoving velocity, and $T_{\gamma}$ and  $T_{\rm S}$ are the CMB brightness temperature and the spin temperature of the neutral hydrogen gas, respectively. All these quantities are evaluated at redshift $z={\nu_{0}}/{\nu}-1$, with $\nu_{0} = 1420.4\ {\rm MHz}$ being the rest-frame frequency of the 21-cm line. During the late EoR that we focus on here, the X-rays emitted by the early X-ray binaries have probably heated the IGM temperature to a level much higher than the CMB temperature \citep{Pritchard2007,Pritchard2012}, and in combination of the efficient Ly$\alpha$ coupling between the spin temperature and the gas kinetic temperature, ensure that $T_{\rm S} \gg T_{\gamma}$. We will assume $T_{\rm S} \gg T_{\gamma}$ in the following.

We compute the 21-cm power spectrum with statistical errors for the three SSA models, and compare them at the same neutral fractions in Fig.~\ref{fig:power-spectrum}. The statistical errors are computed by $<\delta T_b>^2 \Delta^2_{\rm 21}(k)/\sqrt{N}$, where $N$ is the number of independent modes in each $k$-bin. On the small-scale end, as the reionization proceeds to the end, the neutral fibers fragments into islands, and the bubbles-in-island effect becomes less and less prominent, and the small-scale fluctuations tend to be suppressed for all the three models. However, the models differ significantly on large scales. Comparing the colored curves for the three models with the same source parameters of $\{\xi=10$,\  $T^{\rm vir}_{\rm min}=10^{4}\, \rm K$\}, it is obviously seen that the large-scale 21-cm power spectrum ($k< 0.1\,\rm{Mpc^{-1}}$) decreases faster in models with more abundant SSAs, which is in consistent with what was found in \citet{Shukla2016}. In the very late stage ($\bar{x}_{\textsc{H\,i}}\lesssim0.10$), the {\tt islandFAST-noSSA} model predicts the highest fluctuations of 21-cm signal on large scales, because of the larger and fewer islands as shown in the upper panels in Fig.~\ref{fig:xHI-slices}. The more abundant SSAs, or the lower level of the ionizing background, result in the larger number of small islands, significantly suppressing the large-scale power in the 21-cm fluctuations, which could potentially be tested with the upcoming 21-cm power spectrum measurements. The black dot-dashed lines in Fig.~\ref{fig:power-spectrum} shows the 21-cm power spectrum for the {\tt islandFAST-RS} model with $\{\xi=30,\ T^{\rm vir}_{\rm min}=5\times10^{4}\rm K\}$. As compared with the purple dot-dashed lines for the same model but different source parameters, they have a higher amplitude on large scales at all redshifts because of the larger islands resulted from the massive ionizing sources with higher ionizing efficiencies. Similar to the situation in the island size distribution, the effect of SSAs is degenerate with the effect of  source properties for a single redshift, but the degeneracy can be broken by observing the evolution of the 21-cm power spectrum on large scales.

\section{21-cm Observations}\label{ska}
In the above, we have seen that the SSAs have the effects of prolonging the reionization process, and changing the morphology of the neutral islands during the late stages of reionization, which are reflected in both the size distribution of islands and the 21-cm power spectrum.
The 21-cm power spectrum could potentially be detected with the current generation of low-frequency interferometers, such as the LOw Frequency ARray (LOFAR, \citealt{vanHaarlem2013}), the Murchison Widefield Array (MWA, \citealt{Tingay2013}), 
and the Hydrogen Epoch of Reionization Array (HERA, \citealt{DeBoer2017}), and will be more precisely measured by the upcoming  SKA. The neutral regions may also be directly imaged by the SKA \citep{KoopmansSKA2015}.  Here we focus on the SKA, and investigate the possibilities of distinguishing the different SSA models with the power spectrum and images.

Specifically, we assume the phase one configuration of the low-frequency array of SKA (SKA1-Low), which includes 512 antennae stations, each has an effective diameter of 38 m. We consider the {\it core array} with 224 stations within a radius of $500\,\rm{m}$,
and the {\it inner array} of 278 stations distributed in an area of radius 1700 m.
We will take into account the limited angular and spectral resolution and the expected thermal noise of the array, but assume that the foregrounds and RFI removal and calibrations are perfect, though these effects would depend on the technologies still under development (e.g. \citealt{Barry2016}).

To simulate the slices of the observed $\delta T_{\rm b}$ field, we first smooth the $\delta T_{\rm b}$ mock field to match the telescope resolution. The comoving size of a resolved pixel perpendicular to the line of sight is
\begin{align}
l_{\rm xy} = D_{\rm c}(z)\Delta \theta,
\end{align}
where $D_{\rm c}(z)$ is the comoving angular distance to redshift $z$ and $\Delta \theta \sim 1.22\,\lambda/L_{\rm max}$ is the angular resolution of the array, in which $\lambda$ and $L_{\rm max}$ are the observing wavelength and the maximum baseline, respectively.  
The pixel size along the line of sight is related to the observing bandwidth $B$ by
\begin{align}\label{eq.lz}
l_{\rm z} =\displaystyle{\frac{c\,(1+z)^{2}B}{\nu_{0}H(z)}},
\end{align}
where $c$ is the speed of light. In this study, 
we try $B=0.1\MHz$ and  $B=1\MHz$, corresponding to $l_{\rm z} \sim1.48 \Mpc$ and $l_{\rm z} \sim14.8\Mpc$ at $z=6$ respectively. When observing with the narrow band of 0.1 MHz, we use a single slice of depth 1.67 Mpc directly from the 
simulation box, and when observing with the broad band of 1 MHz, 
we average the adjacent 9 slices to get a two-dimensional slice for each redshift $z$. Then each slice is convolved with a Gaussian window of width $\sigma = l_{\rm xy}/\sqrt{8\ln 2}$.

Next, we add the thermal noise to the smoothed slices. For the core array and the inner array of SKA1-Low considered in this work, we simply assume a uniform $uv$-coverage, then the Gaussian thermal noise inside a resolution element corresponding to a scale
$k_{\perp}$ can be written as \citep{KoopmansSKA2015}:
\begin{align}
\sigma_{\rm T}&=\displaystyle{\frac{k_{\perp}}{2\pi}}\, \big(D_{\rm c}^{2}\times\Omega_{\rm FoV}\big)^{1/2}\, \displaystyle{\frac{T_{\rm sys}}{\sqrt{B\,t_{\rm int}}}}\, \sqrt{\displaystyle{\frac{A_{\rm core}A_{\rm eff}}{A_{\rm coll}^{2}}}},
\end{align}
where $\Omega_{\rm FoV}$ is the field of view of the array, $T_{\rm sys}$ is the system temperature, and $t_{\rm int}$ is the integration time. $A_{\rm eff}=\pi\,(19\,\rm m )^2$ is the effective collecting area of each station, which is related to the field of view by $\Omega_{\rm FoV} = {\lambda}^2/A_{\rm eff}$. Here $A_{\rm core}$ is the area over which the antennae are distributed, and the total collecting area is $A_{\rm coll}$.

For the core array of SKA1-Low, $A_{\rm core}=\pi\,(500\,\rm m )^2$, $A_{\rm coll}=224\times A_{\rm eff}$, and 
$\Delta \theta \sim 6.17\ \rm{arcmin}$ at $z=6$ (corresponding to 15.2 comoving Mpc). For the inner array, we have $A_{\rm core}=\pi\,(1700\,\rm m)^2$, $A_{\rm coll}=278\times A_{\rm eff}$, and
 $\Delta \theta \sim 1.81\ \rm{arcmin}$ at $z=6$ (corresponding to 4.45 comoving Mpc). At the frequency range relevant to the 21-cm signals from the EoR, the system temperature is dominated by the sky brightness temperature, and we assume
$T_{\rm sys}=100 + 400\times(\nu/150\,\rm{MHz})^{-2.55}\,\rm K$ \citep{KoopmansSKA2015}. In the following, we adopt deep surveys with $t_{\rm int} = 1000\,\rm{hr}$.
Then the core array has a noise level of $\sigma_{\rm T} \sim 1.4 \mK$ or $0.43 \mK$ for 
$B=0.1\MHz$ or $1\MHz$ respectively, and the inner array will have $\sigma_{\rm T} \sim 12.7 \mK$ ($B=0.1\MHz$) or 
$4.0 \mK$ ($B=1\MHz$). It is found that the increased angular resolution comes at a price, the inner array has a high level of thermal noise
as compared to the cosmological signal, 
and it would be difficult to extract the intrinsic properties of the neutral islands. 
Therefore, we will use the core array in the following analysis.

\begin{figure*}[!htbp]
\centering
\includegraphics[width=0.9\columnwidth]{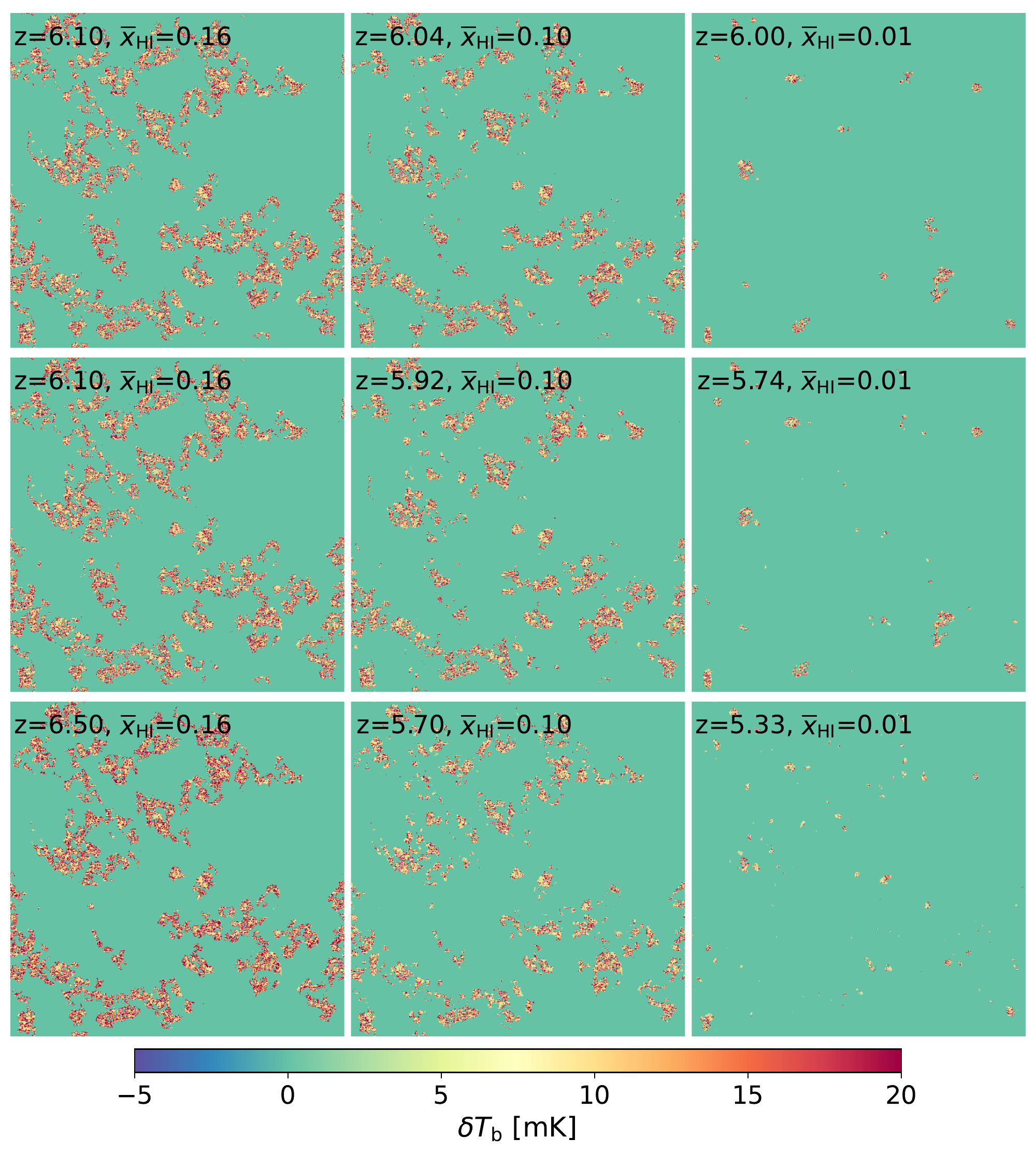}
\includegraphics[width=0.9\columnwidth]{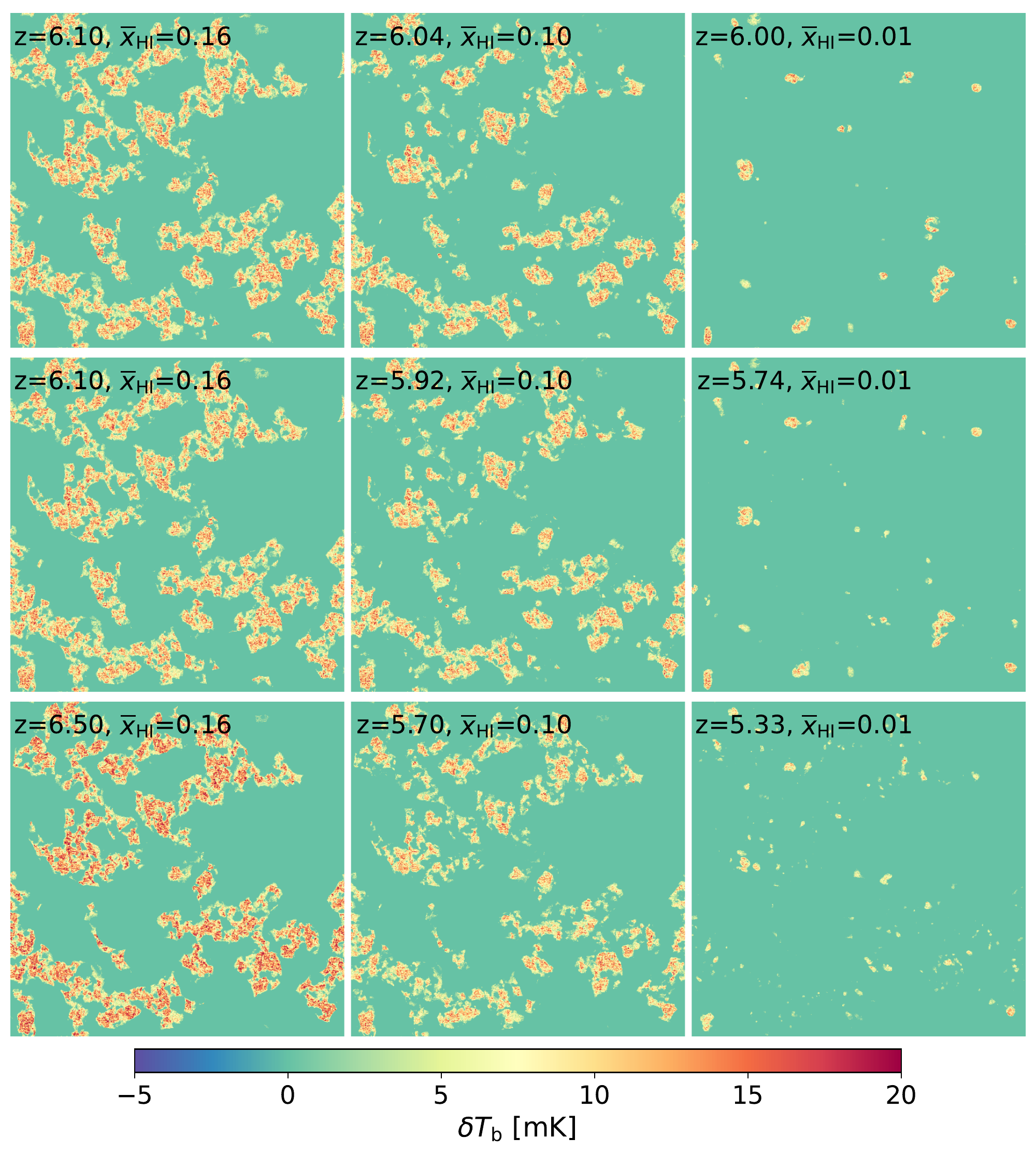}
\includegraphics[width=0.9\columnwidth]{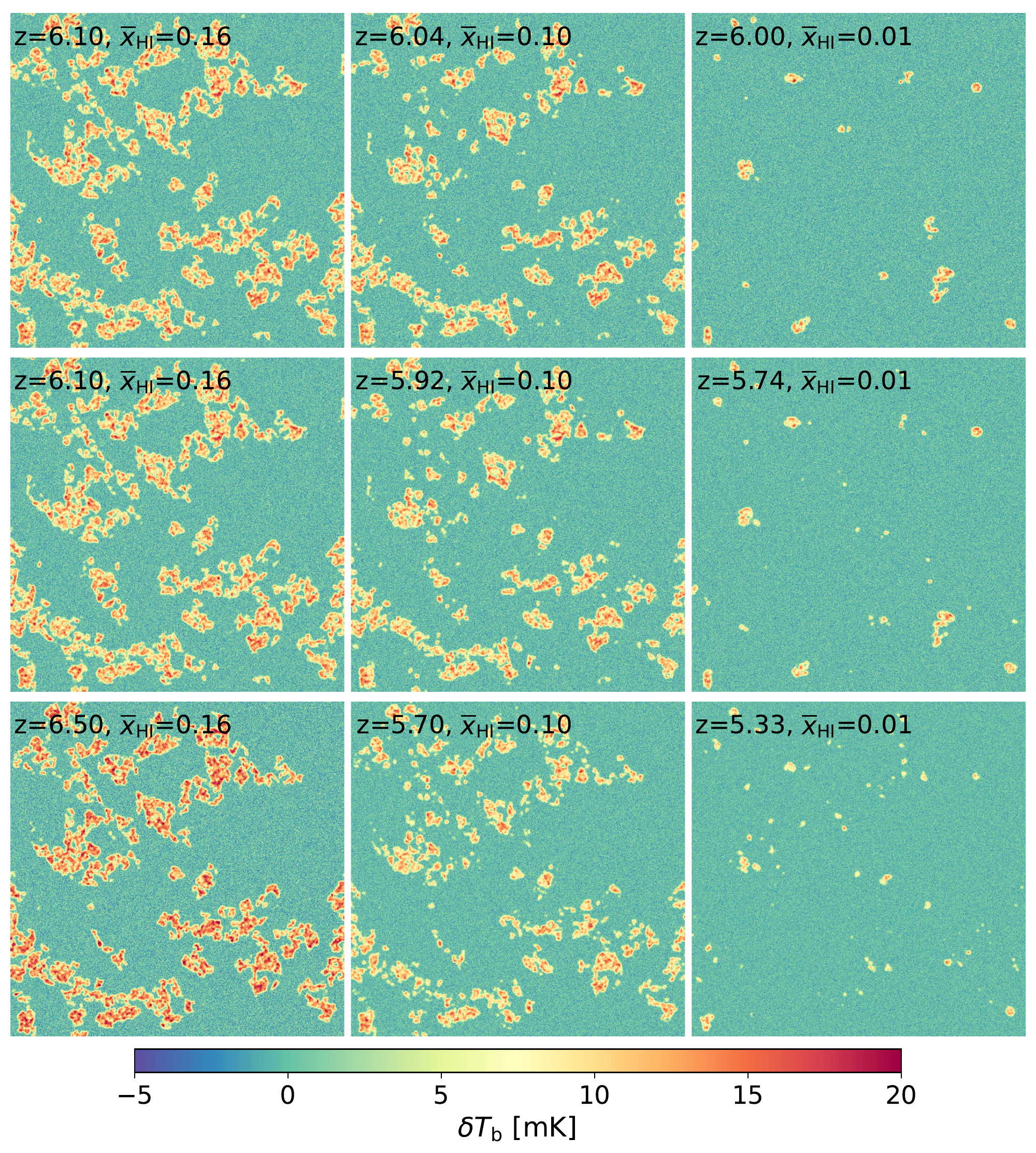}
\includegraphics[width=0.9\columnwidth]{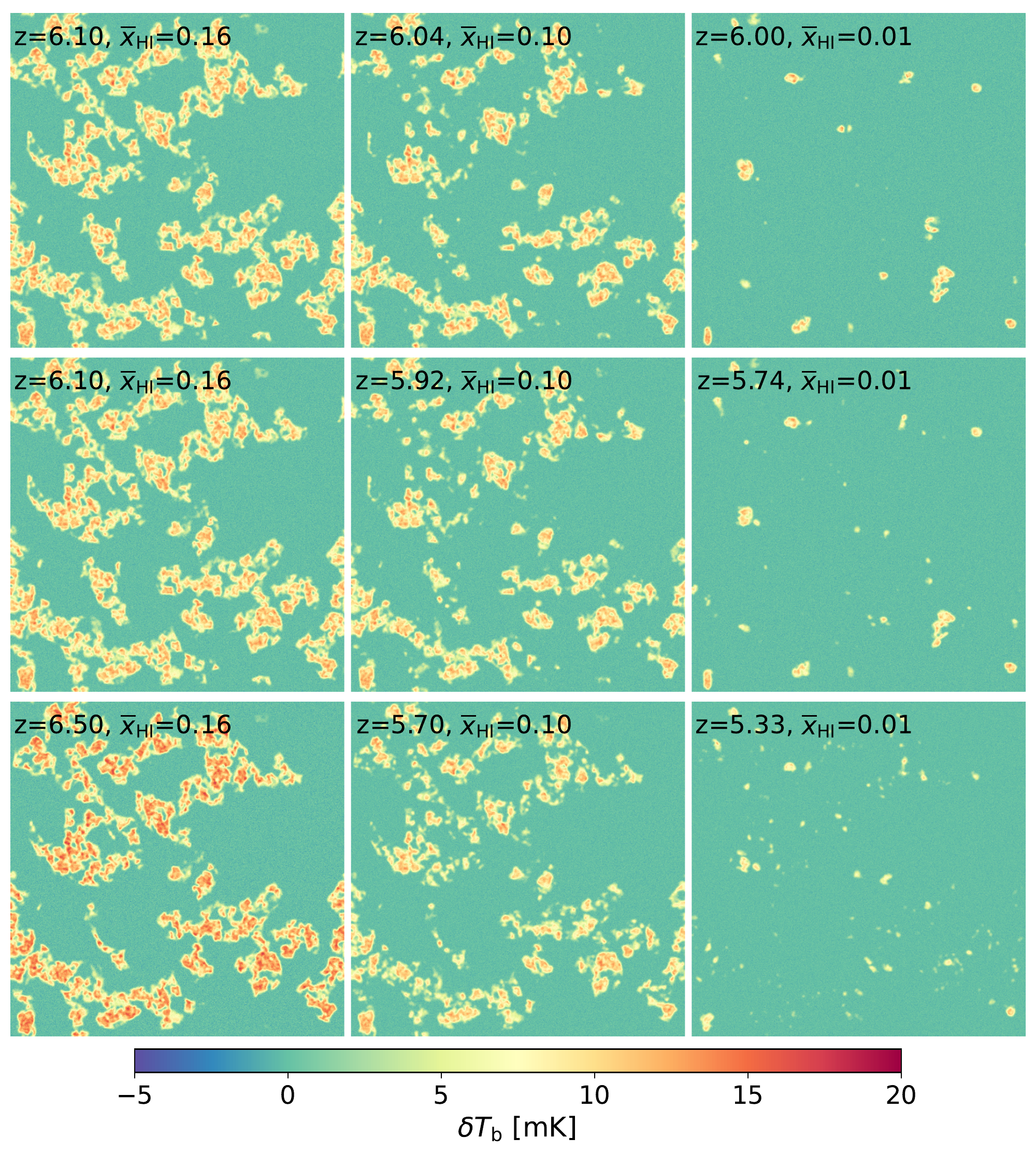}
\caption{The $\delta T_{\rm b}$ slices ({\it upper plots}) and their mock images as observed by the SKA1-Low core array ({\it lower plots}). All slices are $1\, \rm{Gpc}$ on a side in comoving scale. The slices in the left column are $1.67\Mpc$ thick and the observing bandwidth is $0.1\MHz$, and the slices in the right column are $15\,\rm{Mpc}$ thick corresponding to an observing bandwidth of $1\MHz$.
In each plot, the top, middle, and bottom panels are for {\tt islandFAST-noSSA}, {\tt islandFAST-SC}, and {\tt islandFAST-RS} models, respectively, and the three columns correspond to
reionization stages with the mean neutral fractions of $\bar{x}_{\textsc{H\,i}}=0.16$, $0.10$, and $0.01$, from left to right respectively. }
\label{fig:deltaT-slices}
\end{figure*}

Fig.~\ref{fig:deltaT-slices} shows the $\delta T_{\rm b}$ slices from a single slice from the simulation (top-left plot),
 those averaged from 9 adjacent slices (top-right plot), and their corresponding mock images as observed by the SKA1-Low core array using $B=0.1\MHz$ (bottom-left plot) or $B=1\MHz$ (bottom-right plot), respectively.
In each plot, we show the slices predicted by the {\tt islandFAST-noSSA} (top row), {\tt islandFAST-SC} (middle row), and {\tt islandFAST-RS} models (bottom row)  
at fixed neutral fractions of $\bar{x}_{\textsc{H\,i}}=0.16$, $0.10$, and $0.01$ from left to right, with the corresponding redshifts marked on the slices.
The neutral islands are seen in 21-cm emission, while the ionized regions fluctuate around the zero-brightness 
in the mock images due to the thermal noise.

It is seen from the top-left plot that the intrinsic $\delta T_{\rm b}$ slices perfectly follow the ionization field 
(Fig.~\ref{fig:xHI-slices}) as expected, while the averaged $\delta T_{\rm b}$ slices in the top-right plot show clearly the 
projection effect; most small bubbles 
within the islands disappear and the boundaries of islands expand due to the line-of-sight smoothing.
With the limited angular resolution of the SKA1-Low core array (15.2 comoving Mpc in the transverse direction at $z = 6$), 
$\delta T_{\rm b}$ near the boundaries of the islands is lowered, while $\delta T_{\rm b}$ near the edges of ionized regions becomes slightly positive, making the boundaries slurred and enlarging the size of the observed islands, as shown in the bottom plots of Fig.~\ref{fig:deltaT-slices}.

\begin{figure*}[p] 
\centering
\includegraphics[width=0.65\columnwidth]{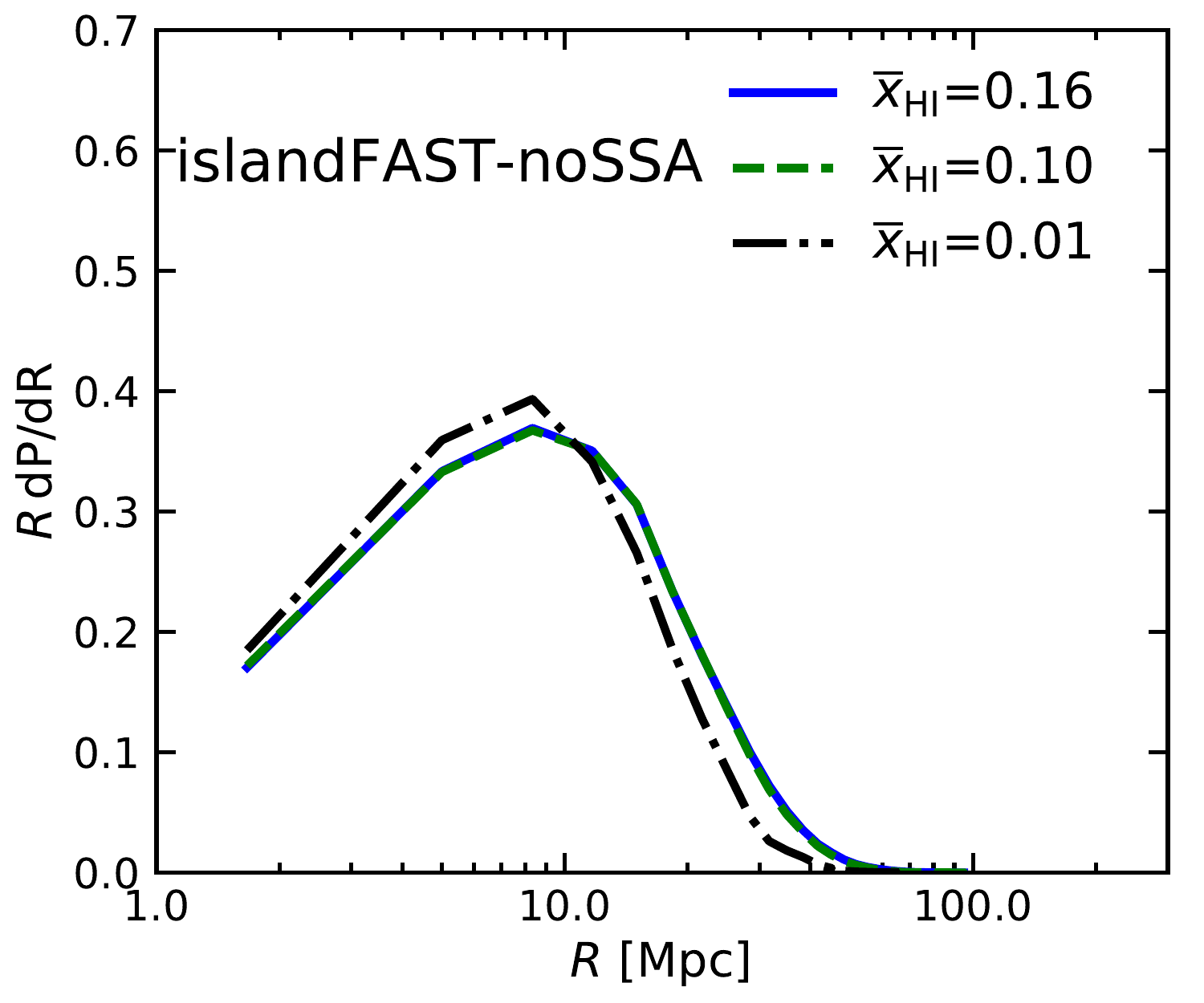}
\includegraphics[width=0.65\columnwidth]{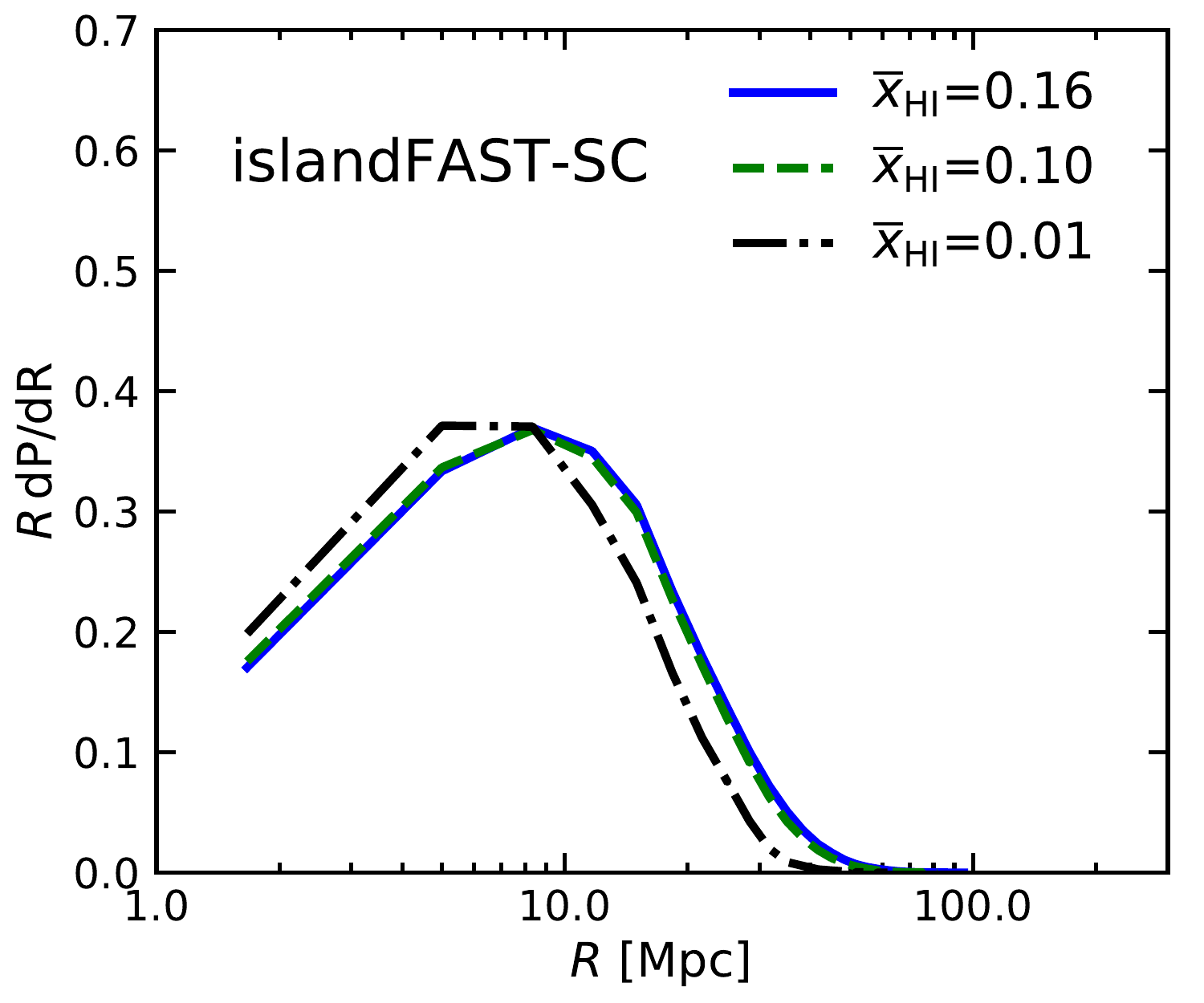}
\includegraphics[width=0.65\columnwidth]{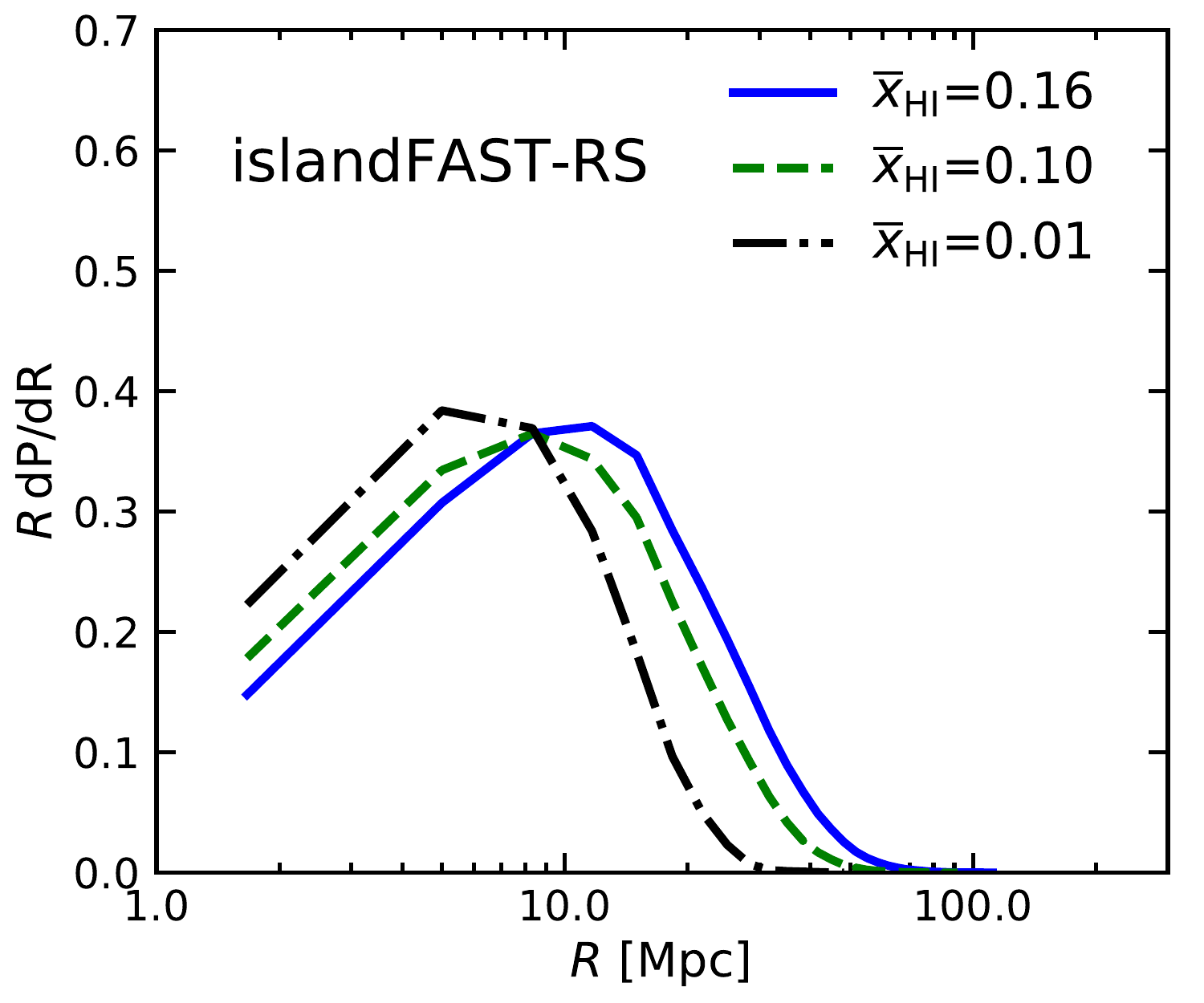}\\
\includegraphics[width=0.65\columnwidth]{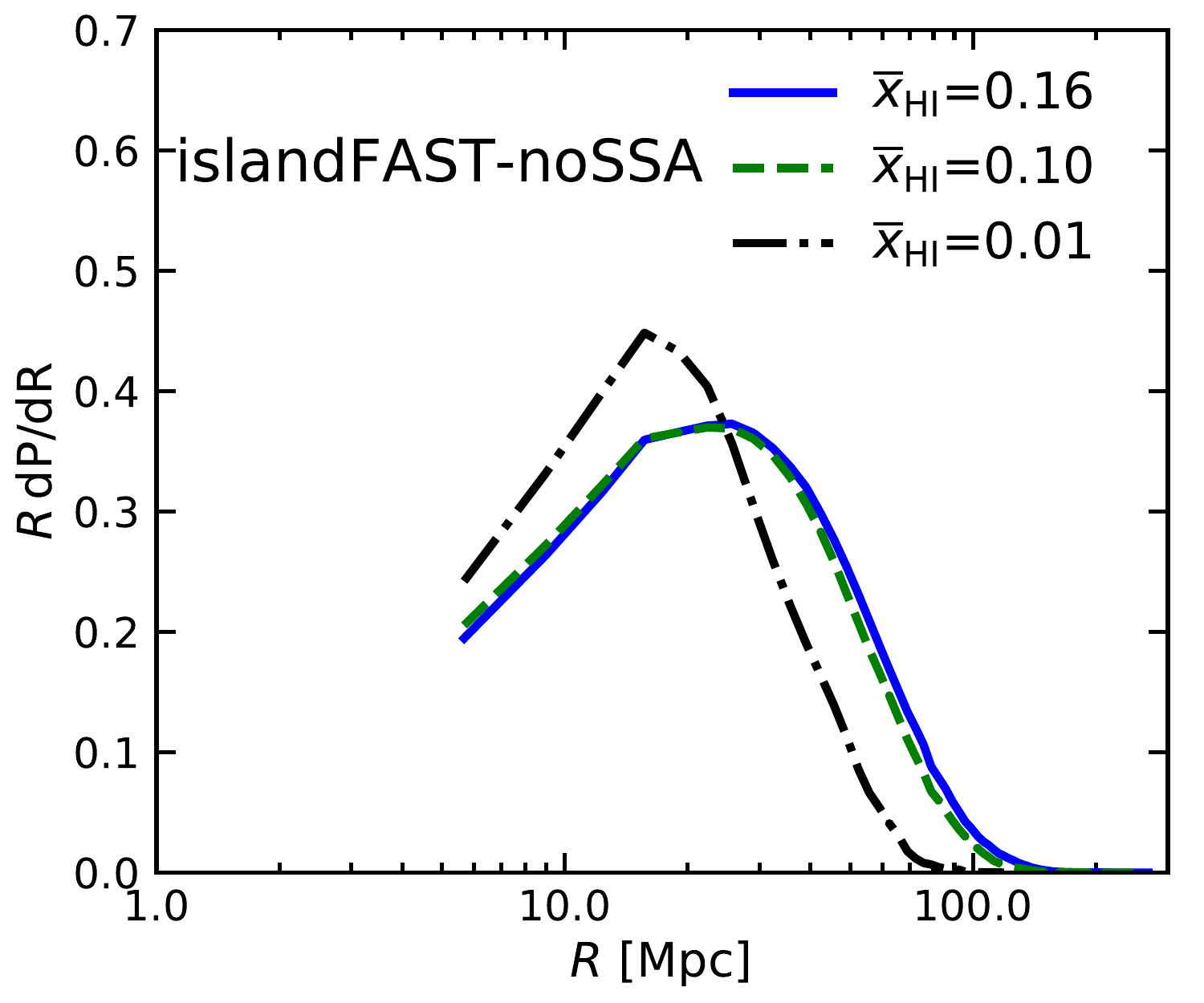}
\includegraphics[width=0.65\columnwidth]{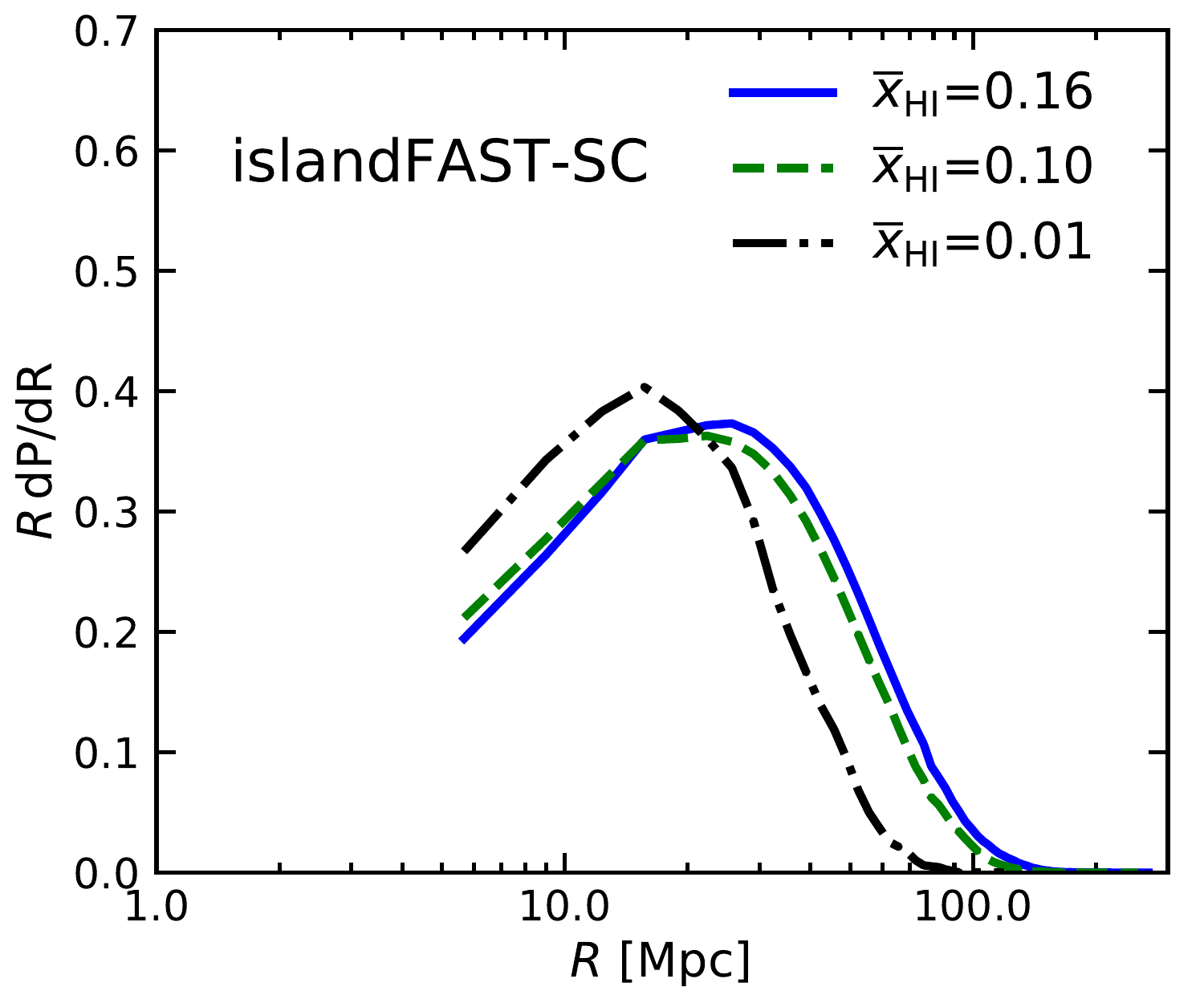}
\includegraphics[width=0.65\columnwidth]{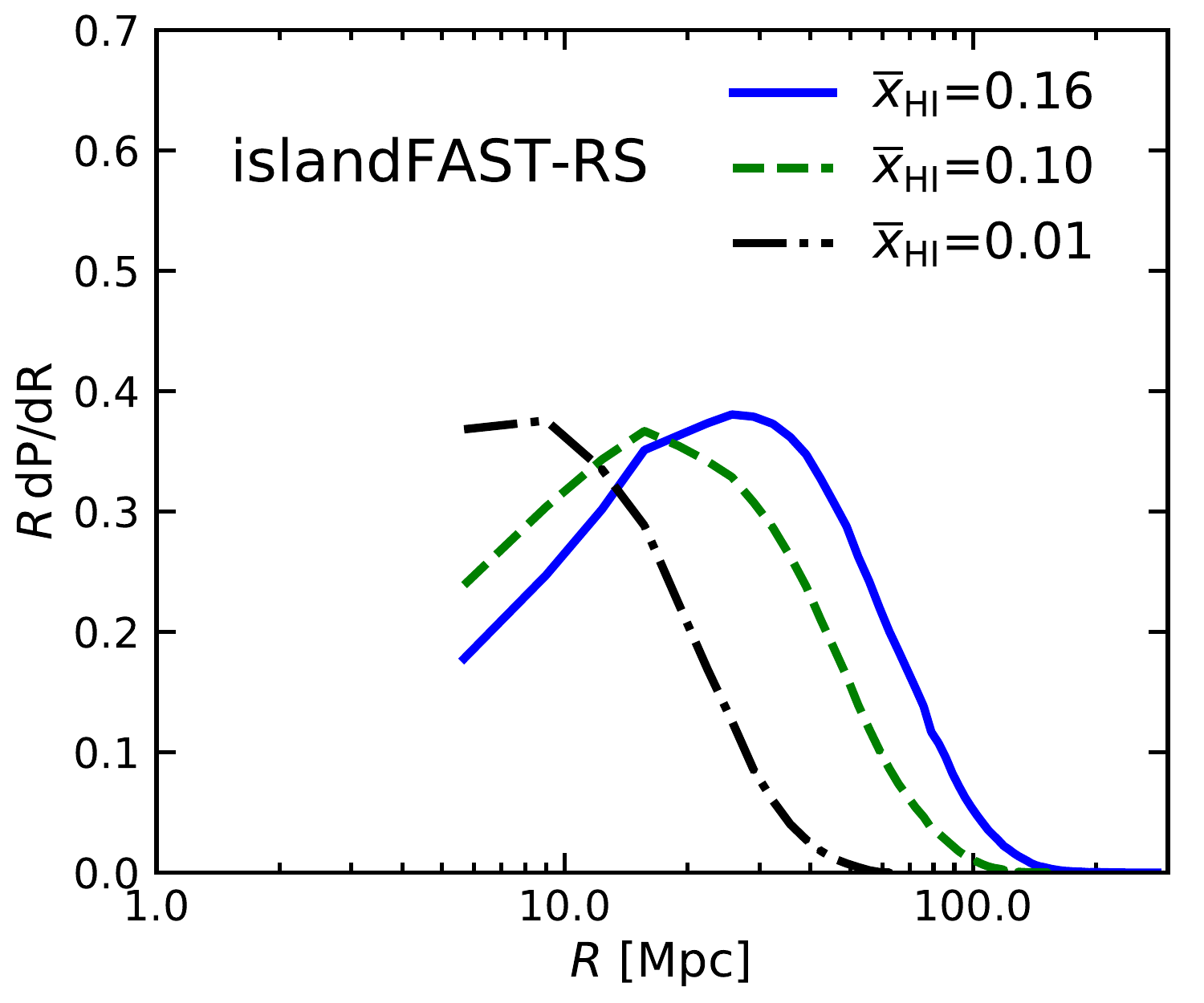}\\
\caption{Size distributions of neutral islands extracted from two-dimensional $\delta T_{\rm b}$ slices from simulations. The upper panels are the results for single slices of depth 1.67 Mpc, and the lower panels are for averaged slices of depth 15 Mpc. In each row, the left, middle, and right panels are from simulations of {\tt islandFAST-noSSA}, {\tt islandFAST-SC}, and {\tt islandFAST-RS}, respectively. In each panel, the blue solid, green dashed, and black dot-dashed curves are for $\bar{x}_{\textsc{H\,i}}=0.16$, $0.10$, and $0.01$, respectively.}
\label{fig:Size-extraction-average}
\end{figure*}

\begin{figure*}[htbp]
\includegraphics[width=0.65\columnwidth]{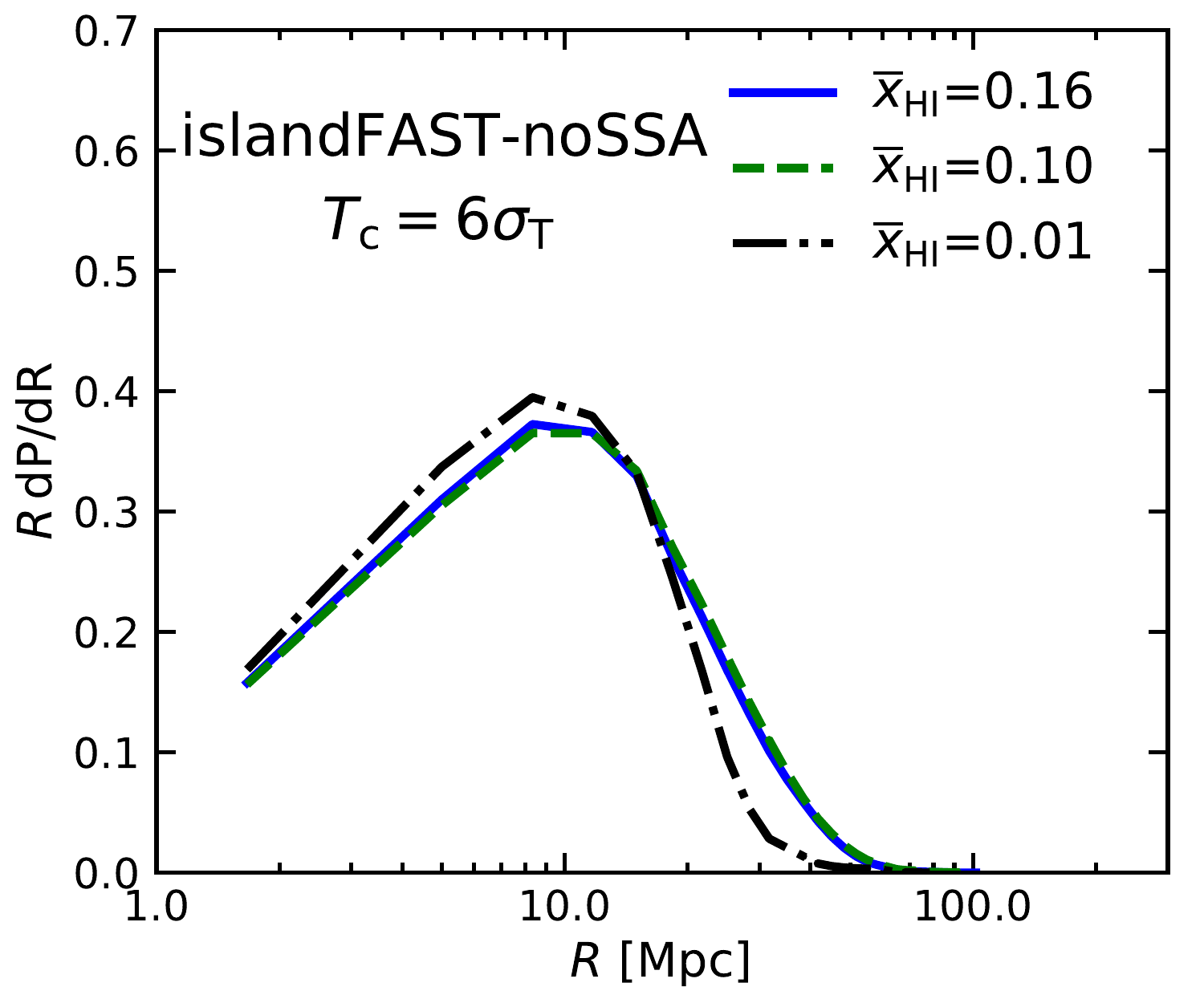}
\includegraphics[width=0.65\columnwidth]{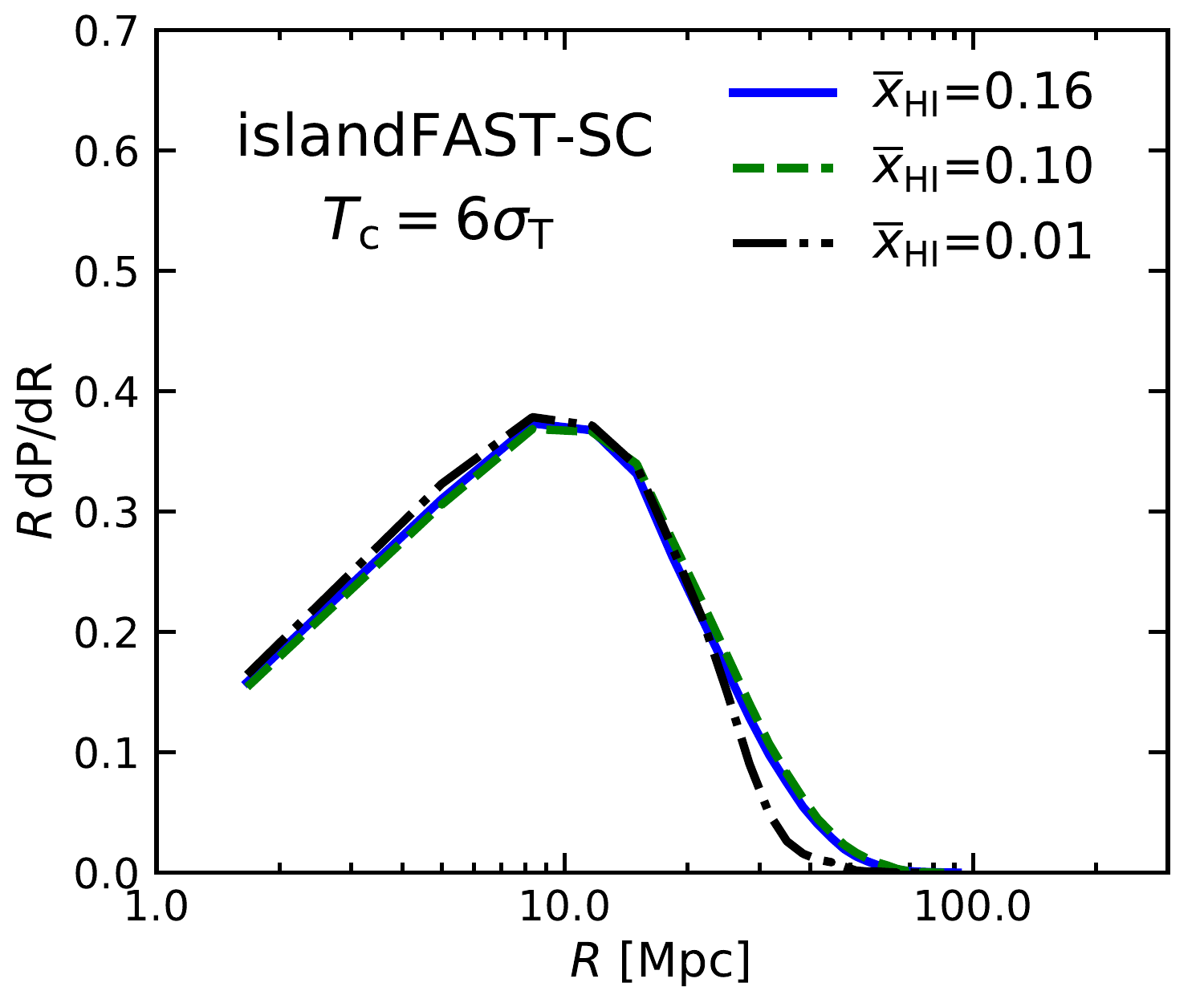}
\includegraphics[width=0.65\columnwidth]{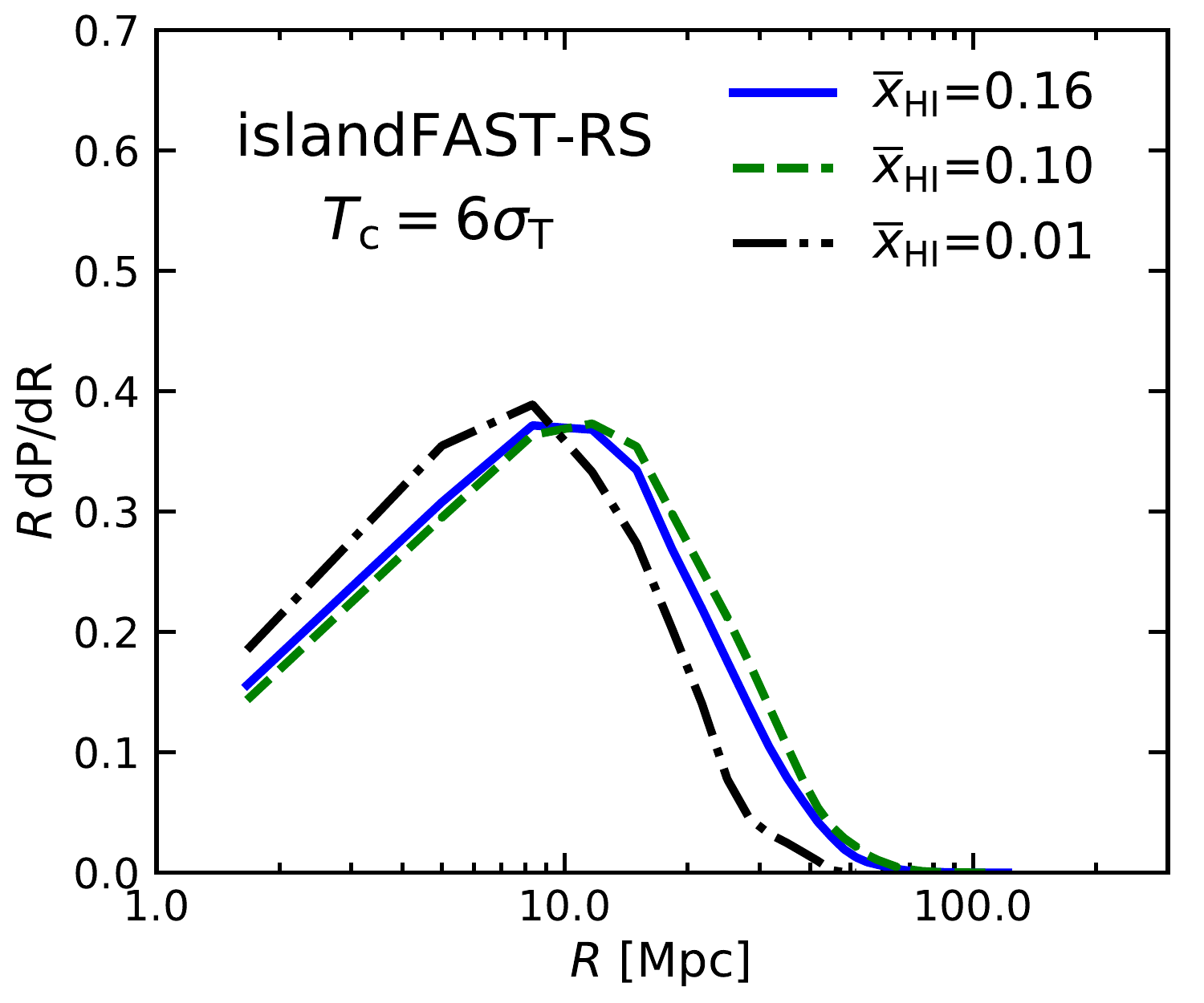}\\
\includegraphics[width=0.65\columnwidth]{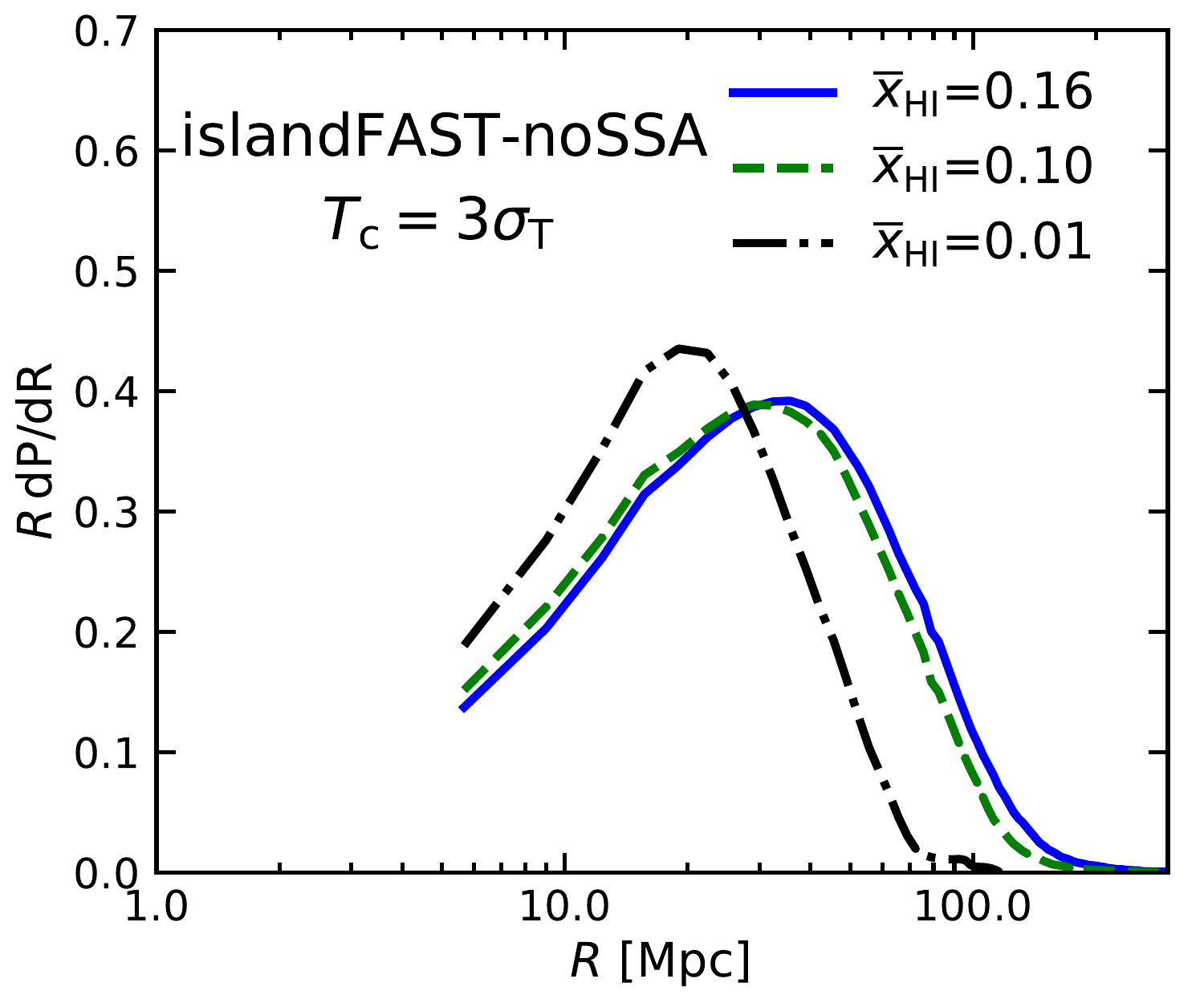}
\includegraphics[width=0.65\columnwidth]{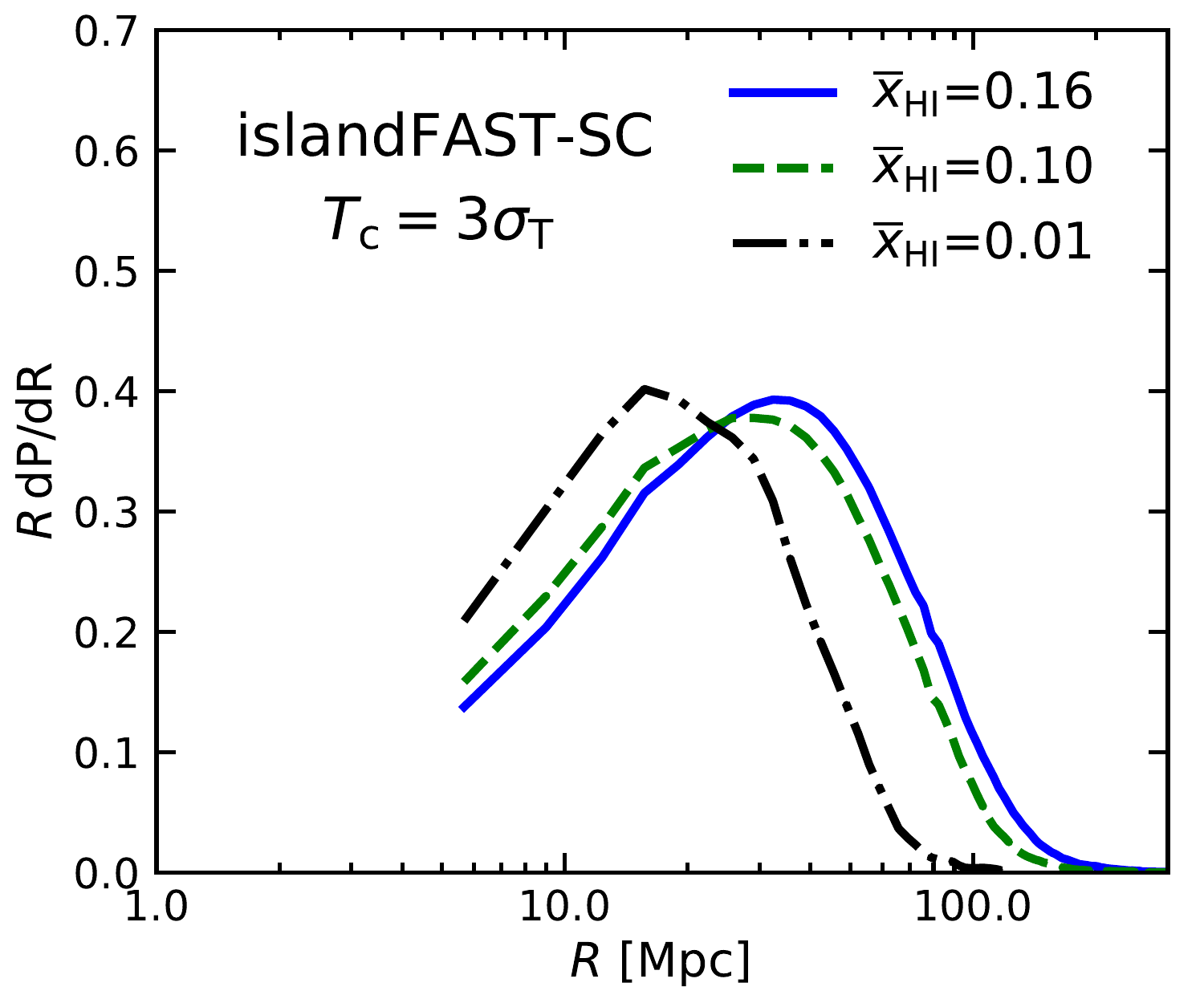}
\includegraphics[width=0.65\columnwidth]{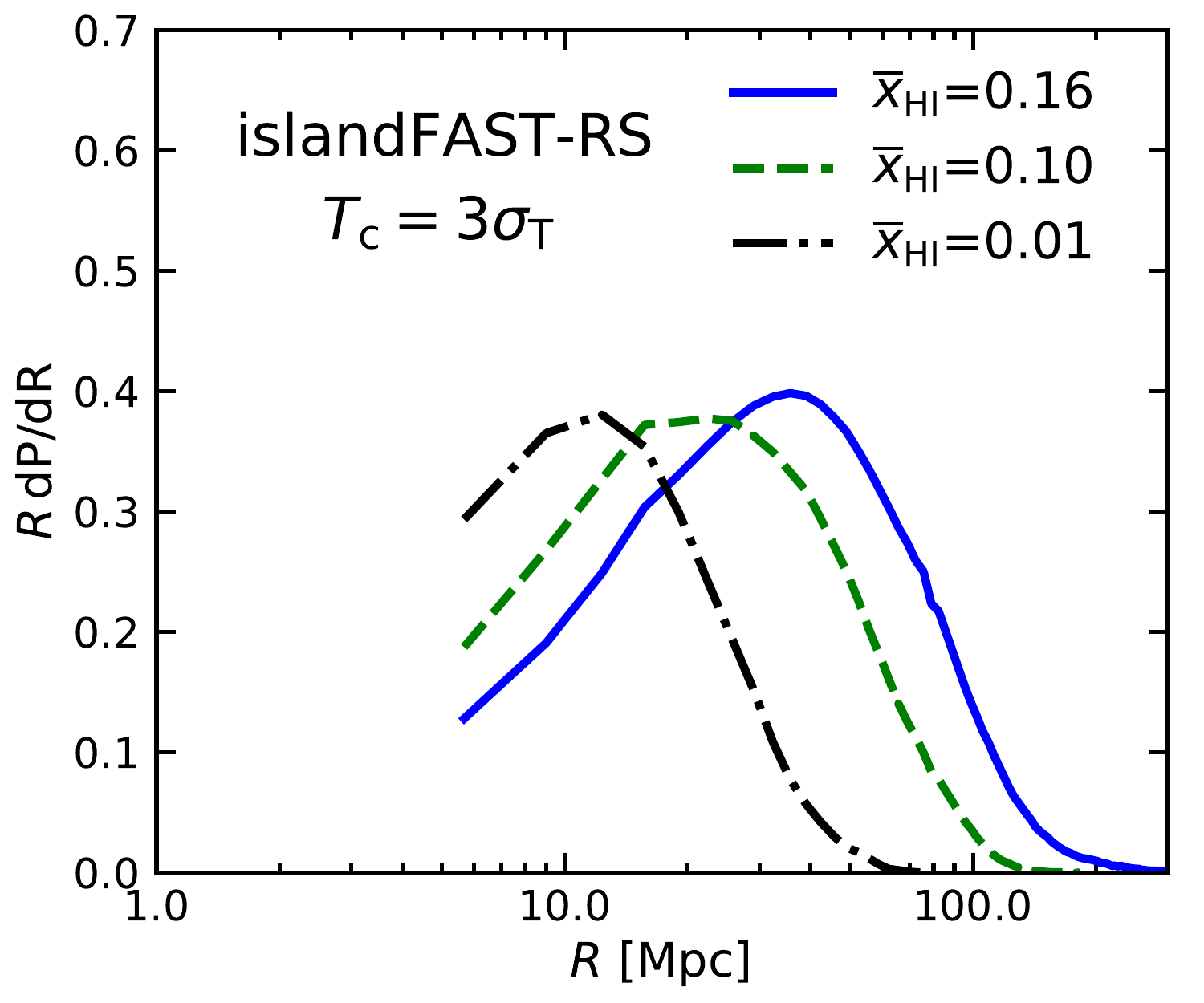}
\centering
\caption{Size distributions of neutral islands extracted from mock $\delta T_{\rm b}$ images as observed by the core array of SKA1-Low, in correspondence to the original $\delta T_{\rm b}$ slices in Fig.~\ref{fig:Size-extraction-average}. The upper panels assume narrow band observations with $B = 0.1\MHz$, and a brightness temperature threshold of $6\,\sigma_{\rm T}$ is used. The lower panels adopt relatively broad band observations with $B = 1\MHz$, and a threshold of $3\,\sigma_{\rm T}$ is used.}
\label{fig:Size-extraction-core}
\end{figure*}

In order to calculate the size distribution of neutral islands that is possibly extractable from two-dimensional $\delta T_{\rm b}$ 
slices, even without any instrumental effects, 
we apply the mean-free-path method to both the single $\delta T_{\rm b}$ slices directly from simulations and the 
averaged slices of depth 15 Mpc, and the results are shown in the upper and lower panels of Fig.~\ref{fig:Size-extraction-average}
respectively.
In each row, the three panels from left to right are for the {\tt islandFAST-noSSA}, {\tt islandFAST-SC}, and {\tt islandFAST-RS} 
models, respectively. We see that the characteristic island scales extracted from single $\delta T_{\rm b}$ slices (upper panels)
can well reflect the typical sizes from the three-dimensional ionization fields as shown in Fig.~\ref{fig:size-models}, which
we refer to as ``intrinsic scale'' below. However, the characteristic scales measured from the averaged $\delta T_{\rm b}$ slices
(lower panels) are much larger than the intrinsic scale. Moreover, when projected onto the two-dimensional $\delta T_{\rm b}$ 
slices, the evolutionary behavior of the size distribution is somewhat distorted. The {\tt islandFAST-noSSA} and {\tt islandFAST-SC} models show an evolutionary trend from $\bar{x}_{\textsc{H\,i}}=0.10$ to $\bar{x}_{\textsc{H\,i}}=0.01$, more obviously in the averaged slices, and this is different from the intrinsic evolutionary behavior in the three-dimensional ionization field. However, 
the evolutionary trend for the different models from $\bar{x}_{\textsc{H\,i}}=0.16$ to $\bar{x}_{\textsc{H\,i}}=0.10$ is retained; 
only the {\tt islandFAST-RS} model shows an obvious evolution in the island scale.

\begin{figure*}[htbp]
\includegraphics[width=0.68\columnwidth]{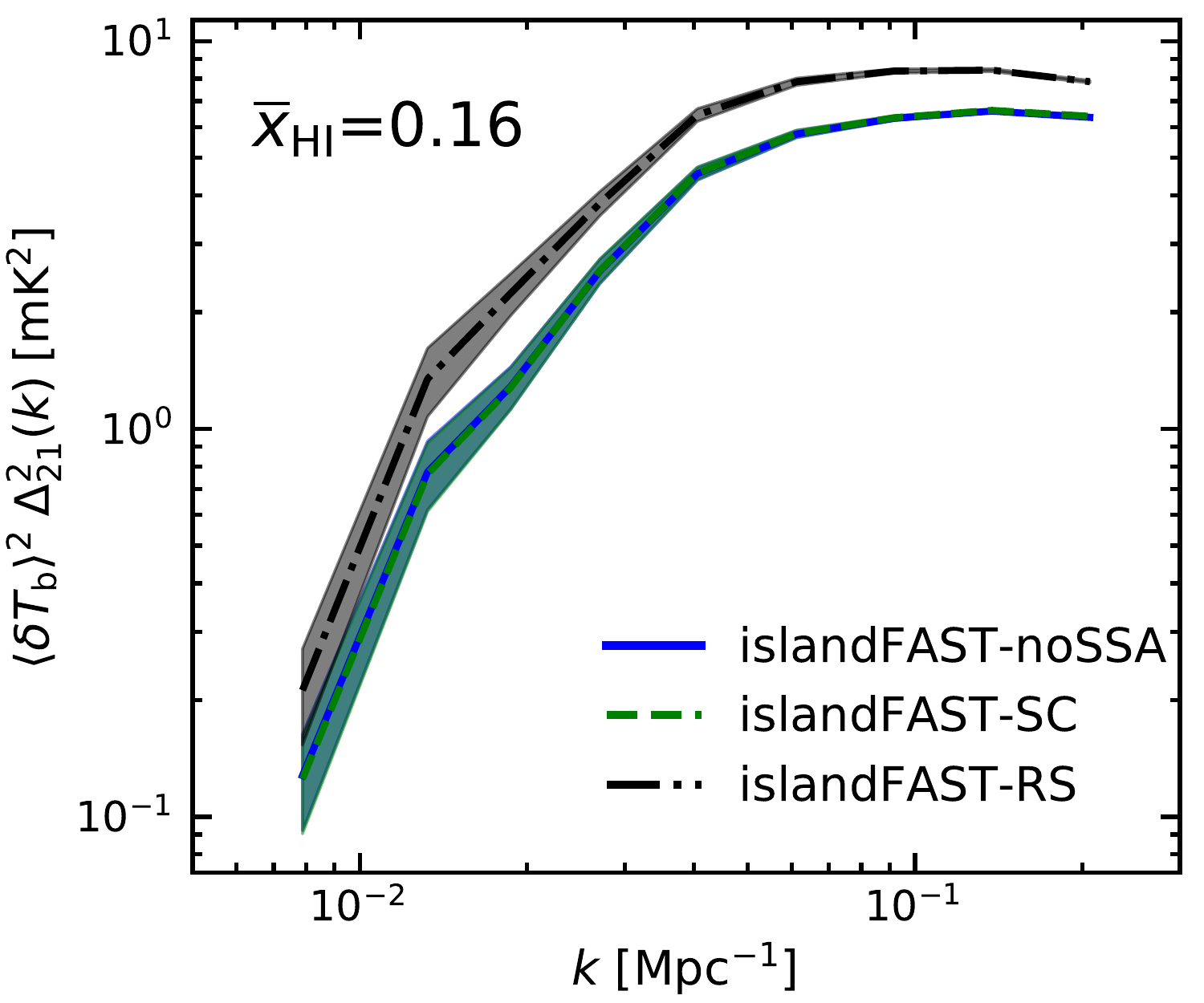}
\includegraphics[width=0.68\columnwidth]{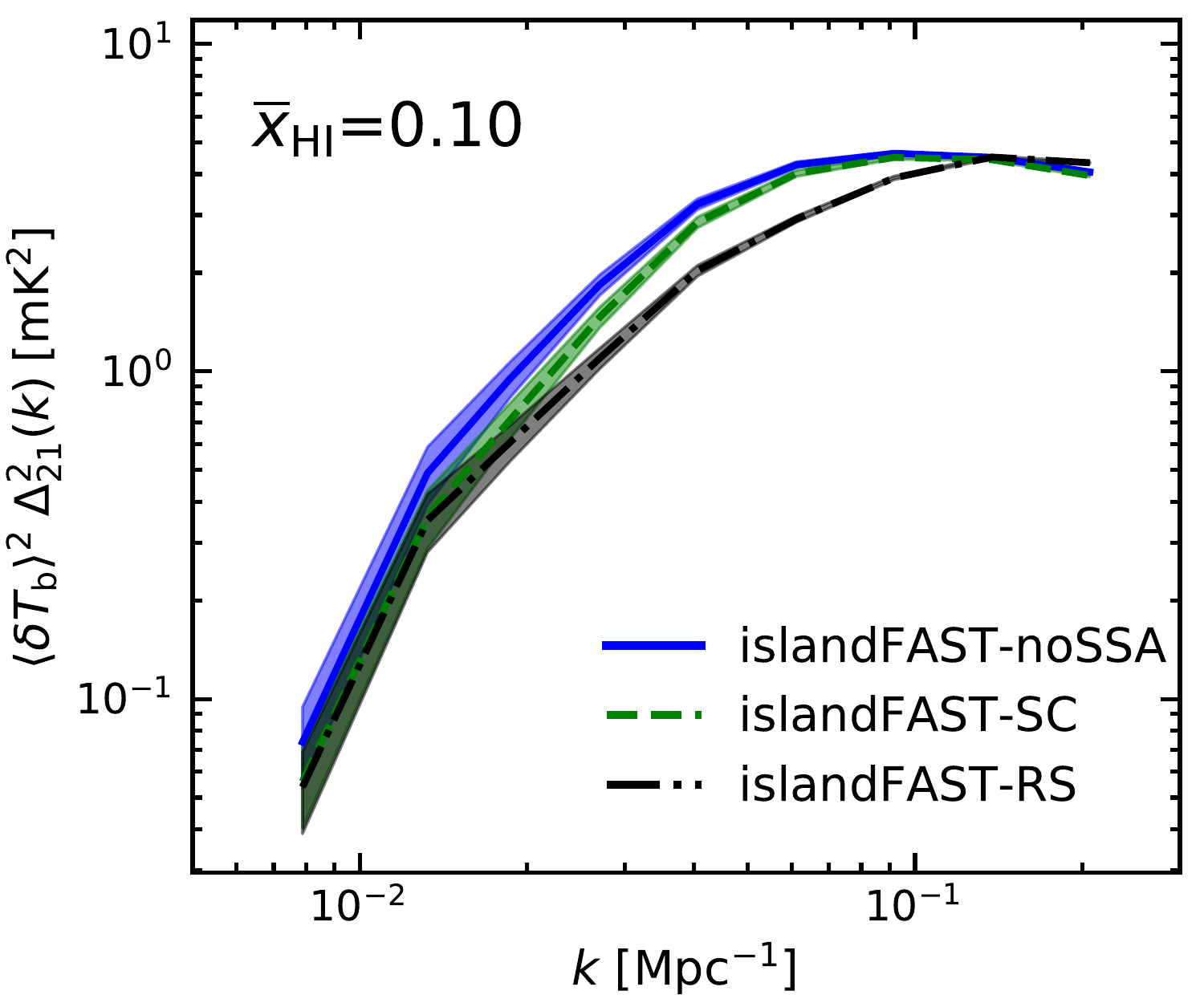}
\includegraphics[width=0.68\columnwidth]{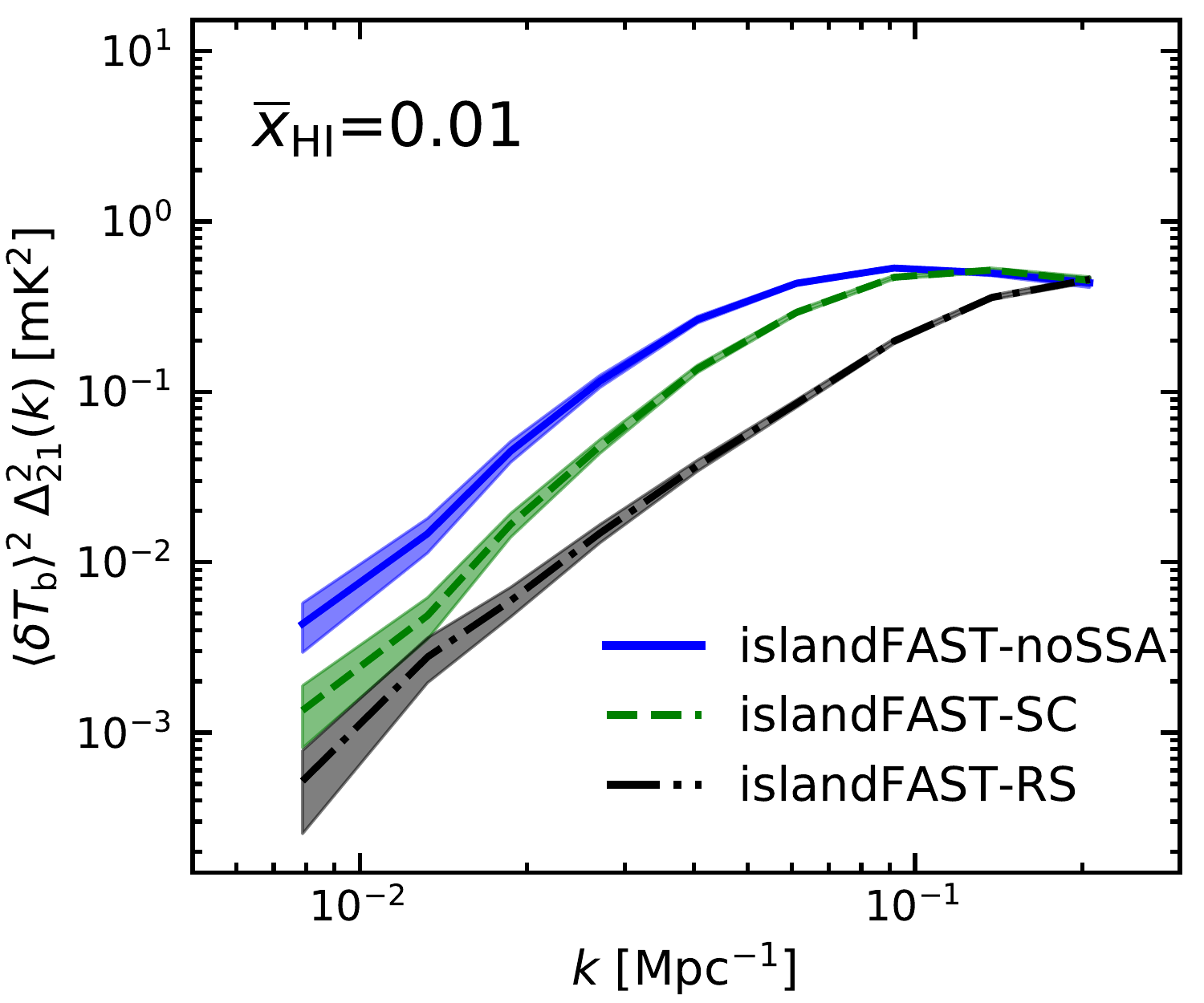}
\centering
\caption{The 21-cm power spectrum at $\bar{x}_{\textsc{H\,i}}=0.16$, 0.10, and 0.01, from left to right respectively.
The blue solid, green dashed, and black dot-dashed lines are for {\tt islandFAST-noSSA}, {\tt islandFAST-SC}, and {\tt islandFAST-RS} simulations, respectively, 
with the shaded regions being the measurement errors expected for the core array of SKA1-Low.}
\label{fig:power-spectrum-SKA}
\end{figure*}

We then apply the mean-free-path algorithm to the mock images as observed by the SKA1-Low core array, 
with a brightness temperature threshold of $T_{\rm c}$. We have experimented with different $T_{\rm c}$ as compared to the 
noise level of $\sigma_{\rm T}$, and found that a relatively high threshold of brightness temperature is required in order to
reconstruct the intrinsic island scales from narrow-band imaging, while a relatively low brightness threshold can be used to extract 
the size distributions of the averaged $\delta T_{\rm b}$ fields from broad-band imaging. Fig.~\ref{fig:Size-extraction-core} shows the size distributions with brightness temperature threshold of $6\, \sigma_{\rm T}$ for narrow-band imaging (upper panels) and those with $3\, \sigma_{\rm T}$ for broad-band imaging (lower panels). It is found that using narrow-band imaging with $B = 0.1\MHz$, we will be able to
extract the intrinsic island scale using a brightness temperature threshold of $6\,\sigma_{\rm T}$. A lower threshold will overestimate the island scales.
Using the broad-band imaging with $B = 1\MHz$, almost all the noises would be eliminated with a threshold value of $3\,\sigma_{\rm T}$. The
reconstructed distributions are very similar to the size distributions shown in the lower panels in Fig.~\ref{fig:Size-extraction-average}, though the extracted characteristic scales are a bit larger. It is clearly that only the {\tt islandFAST-RS} model shows obvious evolution in the extracted island scale throughout the late EoR.

From the above analysis, we see that the line-of-sight and transverse smoothing, and noise jointly affect the observed size distribution and evolution of islands in 21-cm imaging. One needs a balance between the resolution to resolve islands, in both frequency and angular dimensions, and the thermal noise not to overwhelm the signal.
While the smoothing effect can be degenerate with the impact of sparse SSAs, in enlarging the measured island size, 
we can still distinguish the different reionization models by looking for the evolutionary trend from $\bar{x}_{\textsc{H\,i}}=0.16$ to $\bar{x}_{\textsc{H\,i}}=0.10$; only the {\tt islandFAST-RS} model shows an obvious evolution in the measured island scale.
However, the simple mean-free-path method with a single threshold may not extract all useful information about the distributions, and the result may be affected by the choice of threshold. It may require applying multiple thresholds to extract this information. There may also be other more sophisticated and robust ways to extract the bubble or island characteristic scales from observation, e.g. the ``granulometry'' technique \citep{Kakiichi2017}. Such methods should be extensively developed and tested with simulation to ensure correct interpretation of the results.

The different models can also be distinguished with power spectrum measurements. In Section~\ref{21cmps} we noted that the presence of small-scale absorbers mainly affects the large-scale power in the 21-cm brightness temperature fluctuations which is our primary interest. For precise measurement of the large scale power, the core array of SKA1-Low is suitable. Under the assumption of uniform $uv$-coverage, the measurement error on the 21-cm power spectrum due to thermal noise is \citep{KoopmansSKA2015}:
\begin{align}
{\Delta}_{\rm Noise}^{2}&=\displaystyle{\frac{2}{\pi}}\, k^{3/2}\big(D_{\rm c}^{2}\,l_{\rm z}\,\Omega_{\rm FoV}/N_{\rm b}\big)^{1/2}\n\\
& \quad\ \times \bigg(\displaystyle{\frac{T_{\rm sys}}{\sqrt{B\,t_{\rm int}}}}\bigg)^2\, \displaystyle{\frac{A_{\rm core}A_{\rm eff}}{A_{\rm coll}^{2}}},
\end{align}
where $l_{\rm z}$ is the depth of the survey for a bandwidth $B$ given by Eq.~(\ref{eq.lz}), and $N_{\rm b}$ is the number of independent beams. Here we assume $N_{\rm b}=1$, $B=1\,\rm{MHz}$, and $t_{\rm int} = 1000\,\rm{hr}$.
The total error is obtained by adding the statistical error to this thermal noise.

We compute the 21-cm power spectra of the three SSA models for the late EoR from the $\delta T_{\rm b}$-field, smoothed to match the resolution of the core array of SKA1-Low. Fig.~\ref{fig:power-spectrum-SKA} shows the measured 21-cm power spectra for three models, with the total measurement errors indicated by the shaded regions. We find that the differences between the three SSA models at the island stage ($\bar{x}_{\textsc{H\,i}}\lesssim0.10$) are significant with respect to the expected measurement errors for the core array, so at least in principle,  the upcoming SKA1-Low survey should also be able to discriminate the EoR models with power spectrum measurements. In our present case, the abundance of the small scale absorbers and the level of the ionizing background can be distinguished or constrained.

However, we note that the 21cm power spectrum may be biased or distorted by residual foregrounds, though these are not included in the present work. Also, these spectra are quite featureless, so they may also suffer from parameter degeneracies, e.g. the degeneracy between the SSA abundance and the source properties as discussed in Section \ref{21cmps}. The imaging observations which can provide morphological properties of neutral islands and ionized regions are complementary in this aspect.

\section{Conclusion}\label{summary}

In this work, we use a set of semi-numerical simulations to study the effects of the small-scale over-dense absorbers on the large-scale morphology of the under-dense neutral islands during the late EoR. We consider three SSA models, i.e. the {\tt islandFAST-noSSA} model with no SSA and an extremely high ionizing background, the {\tt islandFAST-SC} model with a moderate number density of SSAs and a relatively high ionizing background, and the {\tt islandFAST-RS} model with the most abundant SSAs and a corresponding low level of ionizing background. In agreement with previous works (e.g. \citealt{Alvarez2012,Shukla2016}), we find that the presence of SSAs prolongs the reionization process, and affects the morphology of the ionization field which results in the suppression of the large-scale power in the 21-cm power spectrum.

The morphology of the islands reflects the competitive roles played by the ionizing photons generated inside the islands and those coming from outside.
In the {\tt islandFAST-noSSA} and {\tt islandFAST-SC} models, the ionization is dominated by the photons from outside, i.e. the ionizing background. Small islands could hardly survive, hence when the Universe enters the neutral fibers stage, the evolution of the typical island scale is very slow and stalls at $\sim 10$ comoving Mpc. In the case of {\tt islandFAST-RS} model where the SSA abundance is high and the ionizing background is weak, sources from inside dominates the reionization, then the neutral islands will more easily break into small ones. The size distribution then evolves, and the typical size of the islands is smaller.

Although the characteristic island scale depends on the properties of the ionizing sources, i.e. more massive sources with higher ionizing efficiencies lead to larger neutral islands (see also \citealt{Giri2019}), the evolution in island size distribution and its dependence on the SSA abundance are quite robust. The island size evolution provides a stringent constraint on the SSA abundance, the MFP and the level of the ionizing background during the late EoR. These are key parameters linking the properties of the IGM and the ionizing sources \citep{Becker2021}. We also discussed the extremely long troughs seen in the quasar absorption spectrum which lasts $\sim150$ comoving Mpc at $z=5.5$ \citep{Becker2015}. The probability of finding such long troughs completely driven by large neutral islands in the island stage is quite low for all three SSA models. Either the completion of the reionization is very late, or the reionization history has a large spatial variation (e.g. \citealt{2020ApJ...904..144J,2021ApJ...908...96P}), or much of the space is only partially ionized.

Finally, we investigated the 21-cm tomographic observations with the upcoming SKA1-Low. 
The core array of SKA1-Low has a good sensitivity for both the power spectrum measurement and 21-cm imaging. 
At least in the ideal case, with the foreground perfectly removed, the models with different SSA abundances can be well distinguished with the 21-cm power spectrum measurements at multiple redshifts. 
We show that the evolutionary trend of the neutral islands could be extracted from the images by applying the simple mean-free-path algorithm and choosing a proper threshold of the 21-cm brightness temperature, 
 which then could be used to constrain models of different SSA abundances. However, more sophisticated methods may be needed to extract the full information contained in the image. We conclude that both 21-cm imaging and 21-cm power spectrum can be used to explore the properties of neutral islands.

 In the current study, the effects of SSAs are only incorporated in the late EoR, when the ionization field switches to the neutral island topology. As the SSAs are mainly LLSs in ionized regions, they are important only when the MFP limited by the ionizing fronts is larger than that limited by the SSAs, i.e. $\lambda_{\rm I} \gtrsim \lambda_{\rm abs}$, and the propagation of ionizing photons is effectively limited by the SSAs. During the earlier stages, the propagation of ionizing photons is mainly limited by the ionizing fronts. Therefore, we expect that the presence of SSAs at earlier epochs would not change our results significantly. However, more accurate results require an implementation of the SSA effects throughout the reionization process. We plan to implement a more self-consistent treatment of the SSAs for the whole process in a following work.

\section*{Acknowledgments}
We thank Yue Shao and Jing-Zhao Qi for helpful discussions. This work is supported by the National SKA Program of China (grant No. 2020SKA0110401), the National Natural Science Foundation of China (grant Nos. 11973047, 11975072, and 11633004), the MoST-BRICS Flagship Project (grant No. 2018YFE0120800), the Liaoning Revitalization Talents Program (grant No. XLYC1905011), the science research grants from the China Manned Space Project with NO. CMS-CSST-2021-B01, and the National 111 Project of China (Grant No. B16009).

\bibliography{ref_islands}{}
\bibliographystyle{aasjournal}

\end{CJK*}
\end{document}